\documentclass[11pt]{article} 
\usepackage{hyperref,bm,amssymb,amsmath,natbib}
\usepackage[utf8]{inputenc} 

\usepackage[a4paper,left=2cm,right=2cm,top=2.5cm,bottom=2.5cm]{geometry}

\usepackage{graphicx} 

\usepackage{booktabs} 
\usepackage{array} 
\usepackage{paralist} 
\usepackage{verbatim} 
\usepackage{subfig} 

\usepackage{fancyhdr} 
\usepackage{sectsty}
\allsectionsfont{\sffamily\mdseries\upshape} 

\usepackage[nottoc,notlof,notlot]{tocbibind} 
\usepackage[titles,subfigure]{tocloft} 

\title{kStatistics: Unbiased Estimates of Joint Cumulant Products from
the Multivariate Faà Di Bruno's Formula}
\author{
   Elvira Di Nardo\thanks{Department of Mathematics, `G.Peano'', University of Turin\, Via Carlo Alberto 10, 10123, \url{https://www.elviradinardo.it}, mailto:elvira.dinardo@unito.it}
  \and
  Giuseppe Guarino\thanks{Università Cattolica del Sacro Cuore, 
Largo Agostino Gemelli 8, 00168, Rome (Italy), mailto:giuseppe.guarino@rete.basilicata.it}
}

\date{} 

\begin{document}

\maketitle

\abstract{%
\href{https://cran.r-project.org/web/packages/kStatistics/index.html}{\tt kStatistics}
is a package in \texttt{R} that serves as a unified framework for
estimating univariate and multivariate cumulants as well as products of
univariate and multivariate cumulants of a random sample, using unbiased
estimators with minimum variance. The main computational machinery of
\href{https://cran.r-project.org/web/packages/kStatistics/index.html}{\tt kStatistics}
is an algorithm for computing multi-index partitions. The same algorithm
underlies the general-purpose multivariate Faà di Bruno's formula, which
has been therefore included in the last release of the package. This
formula gives the coefficients of formal power series compositions as
well as the partial derivatives of multivariable function compositions.
One of the most significant applications of this formula is the
possibility to generate many well-known polynomial families as special
cases. So, in the package, there are special functions for generating
very popular polynomial families, such as the Bell polynomials. However
further families can be obtained, for suitable choices of the formal
power series involved in the composition or when suitable symbolic
strategies are employed. In both cases, we give examples on how to
modify the \texttt{R} codes of the package to accomplish this task.
Future developments are addressed at the end of the paper.
}

\hypertarget{introduction}{%
\section{Introduction}\label{introduction}}

Joint cumulants are usually employed for measuring interactions among
two or more random variables simultaneously, extending the familiar
notion of covariance to higher orders. More in details, suppose
\(\boldsymbol{Y}\) a random vector with moment generating function
\(M_{\boldsymbol{Y}}(\boldsymbol{z}),\) for
\(\boldsymbol{z}=(z_1, \ldots, z_m)\) in a suitable neighborhood of
\(\boldsymbol{0}.\) Thus \(M_{\boldsymbol{Y}}(\boldsymbol{z})\) can be
expressed as \begin{equation}
M_{\boldsymbol{Y}}(\boldsymbol{z}) = \exp\big(K_{\boldsymbol{Y}}(\boldsymbol{z}) \big) 
\label{(1bis)}
\end{equation} where \(K_{\boldsymbol{Y}}(\boldsymbol{z})\) is the
cumulant generating function of \(\boldsymbol{Y}.\) If\footnote{If
  \(\boldsymbol{i}\in {\mathbb N}_0^m\) is a multi-index then we set
  \(\boldsymbol{i}! = i_1! \cdots i_m!\) and
  \(|\boldsymbol{i}|=i_1 + \cdots + i_m.\)}
\(\boldsymbol{i}\in {\mathbb N}_0^m\) and \begin{equation}
M_{\boldsymbol{Y}}(\boldsymbol{z})=1 + \sum_{|\boldsymbol{i}| > 0} \frac{{\mathbb E}[\boldsymbol{Y}^{\boldsymbol{i}}]}{\boldsymbol{i}!} \boldsymbol{z}^i \, \qquad \, K_{\boldsymbol{Y}}(\boldsymbol{z}) = \sum_{|\boldsymbol{i}| > 0} \frac{k_{\boldsymbol{i}}(\boldsymbol{Y})}{\boldsymbol{i}!} \boldsymbol{z}^{\boldsymbol{i}}
\label{(2)}
\end{equation} then \(\{k_{\boldsymbol{i}}(\boldsymbol{Y})\}\) are said
the joint cumulants of
\(\{{\mathbb E}[\boldsymbol{Y}^{\boldsymbol{i}}]\}.\) From a theoretical
point of view, cumulants are a useful sequence due to the following
properties \citep{MR3437172}:

\begin{itemize}
\item
  \emph{Orthogonality:} Joint cumulants of independent random vectors
  are zero, that is \(k_{\boldsymbol{i}}(\boldsymbol{Y}) = 0\) for
  \(|\boldsymbol{i}| > 0\) if
  \(\boldsymbol{Y} = (\boldsymbol{Y}_1, \boldsymbol{Y}_2)\) with
  \(\boldsymbol{Y}_1\) independent of \(\boldsymbol{Y}_2.\)
\item
  \emph{Additivity:} Cumulants linearize on independent random vectors,
  that is\\
  \(k_{\boldsymbol{i}}(\boldsymbol{Y}_1 + \boldsymbol{Y}_2) = k_{\boldsymbol{i}}(\boldsymbol{Y}_1) + k_{\boldsymbol{i}}(\boldsymbol{Y}_2)\)
  for \(|\boldsymbol{i}|> 0\) with \(\boldsymbol{Y}_1\) independent of
  \(\boldsymbol{Y}_2.\)
\item
  \emph{Multilinearity:}
  \(k_{\boldsymbol{i}}(A \boldsymbol{Y}) = \sum_{j_1, \ldots, j_m} (A)_{\scriptscriptstyle{i_1}}^{\scriptscriptstyle{j_1}} \cdots (A)_{\scriptscriptstyle{i_m}}^{\scriptscriptstyle{j_m}} k_{\boldsymbol{j}}(\boldsymbol{Y})\)
  for \(|\boldsymbol{i}|>0\) with
  \(A \in {\mathbb R}^m \times {\mathbb R}^m.\)
\item
  \emph{Semi-invariance:} If \(\boldsymbol{b} \in {\mathbb R}^m\) then~
  \(k_{\boldsymbol{i}}(\boldsymbol{Y} + \boldsymbol{b}) = k_{\boldsymbol{i}}(\boldsymbol{Y})\)
  for \(|\boldsymbol{i}| \geq 2\).
\end{itemize}

Thanks to all these properties, joint cumulants have a wide range of
applications: from statistical inference and time series
\citep{MR2303087} to asymptotic theory \citep{rao1999some}, from spatial
statistics modeling \citep{MR2576676} to signal processing
\citep{1458151}, from non-linear systems identification
\citep{systems9020046} to Wiener chaos \citep{Peccati}, just to mention
a few. Indeed it is also well known that cumulants of order greater than
two are zero for random vectors which are Gaussian. Therefore, higher
order cumulants are often used in testing for multivariate Gaussianity
\citep{MR2303087}.

The \(\boldsymbol{i}\)-th multivariate \(k\)-statistic is a symmetric
function of the multivariate random sample whose expectation is the
joint cumulant of order \(\boldsymbol{i}\) of the population characters.
These estimators have minimum variance when compared to all other
unbiased estimators and are built by free-distribution methods without
using sample moments. Due to the properties of joint cumulants,
multivariate \(k\)-statistics are employed to check multivariate
gaussianity \citep{ferreira1997cumulants} or to quantify high-order
interactions among data \citep{6100593}, for applications in topology
inference \citep{9754714}, in neuronal science \citep{MR2721349} and in
mathematical finance \citep{MR4129101}. Polykays are unbiased estimators
of cumulant products \citep{robson1957applications} and are particularly
useful in estimating covariances between \(k\)-statistics
\citep{MR907286}. In the
\href{https://cran.r-project.org/web/packages/kStatistics/index.html}{\tt kStatistics}
package \citep{packkS1}, the
\href{https://www.rdocumentation.org/packages/kStatistics/versions/2.0/topics/nPolyk}{\tt nPolyk}
function provides \(k\)-statistics and polykays as well as their
multivariate generalizations. Further implementations are in
\texttt{Phyton} \citep{smith2020tutorial}, in \texttt{Maple} \citep{DG}
and in \texttt{Mathematica} \citep{MR1890491}.

All these estimators are described with a wealth of details by
\citet{MR1280717} and \citet{MR907286} and their construction relied on
some well-known change of bases in the ring of symmetric polynomials. In
\citet{MR3437172} a different approach is followed using suitable
polynomial families and symbolic strategies. This procedure was the core
of the first release (version 1.0) of the
\href{https://cran.r-project.org/src/contrib/Archive/kStatistics/}{\tt kStatistics}
package \citep{packkS}, as the initial goal was to implement tools for
the estimation of cumulants and cumulant products, both in the
univariate and in the multivariate case. As the referred polynomial
families can be traced back to the generalized (complete exponential)
Bell polynomials, the latest version of the package \citep{packkS1} has
also included procedures to generate these polynomials together with a
number of special cases.

Let us recall that the generalized (complete exponential) Bell
polynomials are a family of polynomials involving multivariable Sheffer
sequences \citep{MR538226}. Among its various applications, we recall
the cumulant polynomial sequences and their connection with special
families of stochastic processes \citep{MR3529335}. Indeed, cumulant
polynomials allow us to compute moments and cumulants of multivariate
L'evy processes \citep{AIP}, subordinated multivariate L'evy processes
\citep{MR4129101} and multivariate compound Poisson processes
\citep{MR3412229}. Further examples can be found in \citet{MR505584},
\citet{MR1915279}, \citet{MR2725499} or \citet{PRIVAULT2021109161}.

The generalized (complete exponential) Bell polynomials arise from the
multivariate Faà di Bruno's formula, whose computation has therefore
been included in the latest version of the
\href{https://cran.r-project.org/web/packages/kStatistics/index.html}{\tt kStatistics}
package. In enumerative combinatorics, Faà di Bruno's formula is
employed in dealing with formal power series. In particular the
multivariate Faà di Bruno's formula gives the \(\boldsymbol{i}\)-th
coefficient of the composition \citep{MR2773373} \begin{equation}
h(\boldsymbol{z}) = f\left(g_1(\boldsymbol{z})-1, \ldots,g_n(\boldsymbol{z})-1\right)
\label{mfaa1}
\end{equation} where \(f\) and \(g_j\) for \(j=1, \ldots, n\) are
(exponential) formal power series \begin{equation} \label{multps1}
f(\boldsymbol{x})= \sum_{|\boldsymbol{t}| \geq 0} f_{\boldsymbol{t}} \frac{\boldsymbol{x}^{\boldsymbol{t}}}{\boldsymbol{t}!}
\quad \hbox{and} \quad 
g_j(\boldsymbol{z}) = \sum_{|\boldsymbol{s}| \geq 0} g_{j; \boldsymbol{s}} \frac{\boldsymbol{z}^{\boldsymbol{s}}}{\boldsymbol{s}!}, 
\end{equation} where
\(\boldsymbol{x}=(x_1, \ldots, x_n),\boldsymbol{z}=(z_1, \ldots, z_m)\)
and\footnote{We use these notations independently if the powers or the
  subscripts are row vectors or column vectors.}
\(\boldsymbol{x}^{\boldsymbol{t}} = x_1^{t_1} \cdots x_n^{t_n},\)
\({\boldsymbol{z}}^{\boldsymbol{s}} = z_1^{s_1} \cdots z_m^{s_m},\)
\(f_{\boldsymbol{t}} = f_{t_1, \ldots, t_n},\)
\(g_{j; \boldsymbol{s}} = g_{j; s_1, \ldots, s_m}\) for
\(j=1,\ldots,n,\) and
\(f_{\boldsymbol{0}}=g_{1; \boldsymbol{0}}= \cdots = g_{n; \boldsymbol{0}}=1.\)
For instance, from \(\eqref{(1bis)}\) and \(\eqref{(2)}\) joint moments
can be recovered from joint cumulants using the multivariate Faà di
Bruno's formula for \(n=1,\)
\(g(\boldsymbol{z}) = 1 + K_{\boldsymbol{Y}}(\boldsymbol{z})\) and
\(f(x)=\exp(x).\) As \(1 + K_{\boldsymbol{Y}}(\boldsymbol{z})=1 +\)
\(\log([M_{\boldsymbol{Y}}(\boldsymbol{z})-1]+1)\) then joint cumulants
can be recovered from joint moments using the multivariate Faà di
Bruno's formula for
\(n=1, g(\boldsymbol{z}) = M_{\boldsymbol{Y}}(\boldsymbol{z})\) and
\(f(x)= 1 + \log (1 + x).\) Let us remark that the exponential form
\(\eqref{multps1}\) of the formal power series \(f\) and \(\{g_j\}\) is
not a constraint. To work with ordinary formal power series, the
multi-index sequence \(\{f_{\boldsymbol{t}}\}\) needs to be replaced by
the sequence \(\{\boldsymbol{t}! f_{\boldsymbol{t}}\}\) as well as the
multi-index sequence \(\{g_{j; \boldsymbol{s}}\}\) by the sequence
\(\{\boldsymbol{s}! g_{j; \boldsymbol{s}}\}\) for \(j=1, \ldots,n.\) In
this case, the multivariate Faà di Bruno's formula gives the coefficient
\({\boldsymbol{i}!} \tilde{h}_{\boldsymbol{i}}\) with
\(\tilde{h}_{\boldsymbol{i}}\) the \({\boldsymbol{i}}\)-th coefficient
of the (ordinary) formal power series composition \(\eqref{mfaa1}\).

The problem of finding suitable and easily manageable expressions of the
multivariate Faà di Bruno's formula has received attention from several
researchers over the years. This is because the multivariate Faà di
Bruno's formula is a very general-purpose tool with many applications.
We refer to the paper of \citet{MR2326318} for a detailed list of
references on this subject and a detailed account of its applications.
Further applications can be found in \citet{MR2301629},
\citet{MR3375628}, \citet{MR3707509} and \citet{PrivaultArxiv}. A
classical way to generate the multivariate Faà di Bruno's formula
involves the partial derivatives of a composition of multivariable
functions. Suppose \(f(\boldsymbol{x})\) and
\(g_1(\boldsymbol{z}), \ldots, g_n(\boldsymbol{z})\) in
\(\eqref{mfaa1}\) be differentiable functions a certain number of times.
The multivariate Faà di Bruno's formula gives the partial derivative of
order \(\boldsymbol{i}\) of \(h(\boldsymbol{z})\) in
\(\boldsymbol{z}_0\) \begin{equation}\label{(hi)}
h_{\boldsymbol{i}} = \frac{\partial^{|\boldsymbol{i}|}}{\partial z_1^{i_1} \cdots
\partial z_m^{i_m}} h(z_1, \ldots, z_m) \!\!\Bigm\lvert_{\boldsymbol{z}=\boldsymbol{z}_0} \qquad \hbox{for $|\boldsymbol{i}|>0,$}
\end{equation} assuming the partial derivatives of order
\(\boldsymbol{t}\) of \(f(\boldsymbol{x})\) exist in
\(\boldsymbol{x}_0=\)
\(\left(g_1(\boldsymbol{z}_0), \ldots,g_n(\boldsymbol{z}_0)\right)\)
\[ \qquad f_{\boldsymbol{t}} = \frac{\partial^{|\boldsymbol{t}|}}{\partial x_1^{t_1} \cdots
\partial x_n^{t_n}} f(x_1, \ldots, x_n) \!\!\Bigm\lvert_{\boldsymbol{x}=\boldsymbol{x}_0}  \qquad \hbox{for $0 < |\boldsymbol{t}| \leq |\boldsymbol{i}|,$}\]
and the partial derivatives of order \(\boldsymbol{s}\) of
\(g_j(\boldsymbol{z})\) exist in \(\boldsymbol{z}_0\) for
\(j=1,\ldots,n\) \[g_{j,\boldsymbol{s}} = 
\frac{\partial^{|\boldsymbol{s}|}}{\partial z_1^{s_1} \cdots
\partial z_m^{s_m}} g_j(z_1, \ldots, z_m) \!\!\Bigm\lvert_{\boldsymbol{z}=\boldsymbol{z}_0} \qquad \hbox{for $0 < |\boldsymbol{s}| \leq |\boldsymbol{i}|$}.\]

There are various ways to express \(h_{\boldsymbol{i}}\) in
\(\eqref{(hi)}\), see for example \citet{MR1781515}, \citet{MR2005276}
and \citet{MR2515761}. Symbolic manipulation using \texttt{Macsyma},
\texttt{Maple}, \texttt{Mathematica}, etc. can produce any required
order of \(\eqref{(hi)}\), by applying the chain rule recursively and
using a function that provides partial derivatives. Also in \texttt{R},
there are some functions for computing partial derivatives
\citep{DERIV}. Despite its conceptual simplicity, applications of the
chain rule become impractical for its cumbersome computation even for
small values of its order. As the number of additive terms becomes huge,
the output is often untidy and further manipulations are required to
simplify the result. By using combinatorial methods, \citet{MR1325915}
have carried out the following expression of the multivariate Faà di
Bruno's formula \begin{equation} \label{multfaa}
h_{\boldsymbol{i}} = \boldsymbol{i}! \sum_{1 \leq |\boldsymbol{t}| \leq |\boldsymbol{i}|}
f_{\boldsymbol{t}} \sum_{k=1}^{|\boldsymbol{i}|} \sum_{p_k(\boldsymbol{i}, \boldsymbol{t})} \prod_{j=1}^k \frac{({\mathfrak g}_{\boldsymbol{l}_j})^{\boldsymbol{q}_j}}{\boldsymbol{q}_j! (\boldsymbol{l}_j!)^{|\boldsymbol{q}_j|}}
\end{equation} where
\(({\mathfrak g}_{\boldsymbol{s}})^{\boldsymbol{q}}=\prod_{j=1}^{n} (g_{j,\boldsymbol{s}})^{q_j}\)
with \({\boldsymbol{q}}=(q_1, \ldots, q_n)\) and \[
p_k(\boldsymbol{i}, \boldsymbol{t}) = \left\{(\boldsymbol{q}_1, \ldots, \boldsymbol{q}_k;
\boldsymbol{l}_1, \ldots, \boldsymbol{l}_k): |\boldsymbol{q}_j|>0, \sum_{j=1}^k \boldsymbol{q}_j = \boldsymbol{t},  \sum_{j=1}^k |\boldsymbol{q}_j|\boldsymbol{l}_j  = \boldsymbol{i}
\right\}\] with
\(\boldsymbol{q}_1, \ldots, \boldsymbol{q}_k \in {\mathbb N}_0^{n}\) and
\(\boldsymbol{l}_1, \ldots, \boldsymbol{l}_k \in {\mathbb N}_0^{m}\)
such that\footnote{If
  \(\boldsymbol{\mu}, \boldsymbol{\nu} \in {\mathbb N}_0^m\) we have
  \(\boldsymbol{\mu} \prec \boldsymbol{\nu}\) if
  \(|\boldsymbol{\mu}| < |\boldsymbol{\nu}|\) or
  \(|\boldsymbol{\mu}| = |\boldsymbol{\nu}|\) and \(\mu_1 < \nu_1\) or
  \(|\boldsymbol{\mu}| = |\boldsymbol{\nu}|\) and
  \(\mu_1 = \nu_1, \ldots, \mu_k = \nu_k, \mu_{k+1} < \nu_{k+1}\) for
  some \(1 \leq k < m.\)}
\(\boldsymbol{0} \prec \boldsymbol{l}_1 \prec \ldots \prec \boldsymbol{l}_k.\)

A completely different approach concerns the combinatorics of partial
derivatives as \citet{MR2200529} pointed out for the
univariate-multivariate composition using multisets and collapsing
partitions. Motivated by his results and using the umbral calculus,
which is a symbolic method particularly useful in dealing with formal
power series \(\eqref{multps1}\), the combinatorics behind
\(\eqref{multfaa}\) has been simplified and a different expression has
been given in \citet{MR2773373}. The key tool is the notion of partition
of a multi-index which parallels the multiset partitions given in
\citet{MR2200529}.

The contribution of this paper is multi-sided. We explain how to recover
in \texttt{R} a multi-index partition, which is a combinatorial device.
For statistical purposes, we show how to recover \(k\)-statistics and
their multivariate generalizations using the referred polynomial
approach and multi-index partitions. Then, we explain the main steps of
the
\href{https://www.rdocumentation.org/packages/kStatistics/versions/2.0/topics/MFB}{\tt MFB}
function producing the multivariate Faà di Bruno's formula, without any
reference to the umbral calculus or chain rules and whose applications
go beyond statistical purposes. The main idea is to expand the
multivariable polynomial \[ 
\sum {\boldsymbol{i} \choose \boldsymbol{s}_1,\ldots,\boldsymbol{s}_n} q_{1,\boldsymbol{s}_1}(y_1) \cdots q_{n,\boldsymbol{s}_n}(y_n)
\] where
\(q_{1,\boldsymbol{s}_1}(y_1) \ldots q_{n,\boldsymbol{s}_n}(y_n)\) are
suitable polynomials and the sum is over all the compositions of
\(\boldsymbol{i}\) in \(n\) parts, that is all the \(m\)-tuples
\((\boldsymbol{s}_1,\ldots,\boldsymbol{s}_n)\) of non-negative integers
such that
\(\boldsymbol{s}_1 + \cdots + \boldsymbol{s}_n = \boldsymbol{i}.\)
Readers interested in the umbral setting may refer to \citet{MR3437172}
and references therein.

Consequently, the
\href{https://www.rdocumentation.org/packages/kStatistics/versions/2.0/topics/MFB}{MFB}
function gives an efficient computation of the following compositions:

\begin{itemize}
\item
  univariate with univariate, that is \(n=m=1;\)
\item
  univariate with multivariate, that is \(n=1\) and \(m >1;\)
\item
  multivariate with univariate, that is \(n >1\) and \(m=1;\)
\item
  multivariate with multivariate, that is \(n >1\) and \(m>1.\)
\end{itemize}

The
\href{https://cran.r-project.org/web/packages/kStatistics/index.html}{\tt kStatistics}
package includes additional functions, for some of the most widespread
applications of the multivariate Faà di Bruno's formula. Indeed, not
only this formula permits to generate joint cumulants and their inverse
relations, but also further general families of polynomials. Therefore,
we have set up special procedures for those families used very often in
applications. These functions should be considered an easy to manage
interfaces of the
\href{https://www.rdocumentation.org/packages/kStatistics/versions/2.0/topics/MFB}{MFB}
function, with the aim of simplifying its application. Moreover, since
the \texttt{R} codes are free, the user might follow similar steps to
generate polynomial families not included in the package but always
coming from the multivariate Faà di Bruno's formula. The construction of
new families of polynomials can be done mainly in two ways. The first
way is to choose appropriately the coefficients
\(\{f_{\boldsymbol{t}}\}\) and \(\{g_{j; \boldsymbol{s}}\}\) in
\(\eqref{multps1}\). The second way is to use some suitable symbolic
strategies, as discussed in \citet{MR3437172}. For both cases, we
provide examples.

The paper is organized as follows. The next section explains the main
steps of the algorithm that produces multi-index partitions with
particular emphasis on its combinatorics. Then we present the symbolic
strategy to generate \(k\)-statistics and their generalizations using
suitable polynomial sequences and multi-index partitions. The subsequent
section deals with generalized (complete exponential) Bell polynomials
and some special cases corresponding to well-known families of
polynomials. We have also included the procedures to generate joint
cumulants from joint moments and viceversa. In the last section we
explain the main steps of the algorithm to produce the multivariate Faà
di Bruno's formula. We give examples of how to build new polynomials not
included in the package. Some concluding remarks end the paper.

\hypertarget{partitions-of-a-multi-index}{%
\section{Partitions of a
multi-index}\label{partitions-of-a-multi-index}}

Most routines of the
\href{https://cran.r-project.org/web/packages/kStatistics/index.html}{\tt kStatistics}
package use the partitions of a multi-index \(\boldsymbol{i}.\)
Therefore, before describing any of these routines, we recall the notion
of multi-index partition and describe the algorithm for its construction
as implemented in the
\href{https://www.rdocumentation.org/packages/kStatistics/versions/2.0/topics/mkmSet}{\tt mkmSet} function of the package.

A partition of the multi-index
\(\boldsymbol{i} = (i_1, \ldots, i_m) \in {\mathbb N}_0^m\) is a matrix
\(\Lambda = (\boldsymbol{\lambda}_1^{r_1}, \boldsymbol{\lambda}_2^{r_2}, \ldots)\)
of non-negative integers with \(m\) rows and no zero columns such that

\begin{itemize}
\item
  \(r_1 \geq 1\) columns are equal to \(\boldsymbol{\lambda}_1,\)
  \(r_2 \geq 1\) columns are equal to \(\boldsymbol{\lambda}_2\) and so
  on;
\item
  the columns
  \(\boldsymbol{\lambda}_1 < \boldsymbol{\lambda}_2 < \ldots\) are in
  lexicographic order\footnote{As example \((a_1,b_1) < (a_2,b_2)\) if
    \(a_1 < a_2\) or \(a_1=a_2\) and \(b_1<b_2.\)};
\item
  the sum of the integers in the \(t\)-th row is equal to \(i_t,\) that
  is \(\lambda_{t 1}+\lambda_{t 2}+\cdots = i_t\) for
  \(t = 1,2,\ldots,m.\)
\end{itemize}

\noindent We write \(\Lambda \vdash \boldsymbol{i}\) to denote that
\(\Lambda\) is a partition of \(\boldsymbol{i}.\) Some further notations
are:

\begin{itemize}
\item
  \(\mathfrak{m}(\Lambda)=(r_1, r_2, \ldots),\) the vector of
  multiplicities of
  \(\boldsymbol{\lambda}_1, \boldsymbol{\lambda}_2, \ldots\)
\item
  \(l(\Lambda)=|\mathfrak{m}(\Lambda)|=r_1 + r_2 + \cdots,\) the number
  of columns of \(\Lambda\) with \(l(\Lambda )=0\) if
  \(\Lambda \vdash \boldsymbol{0}\)
\item
  \(\Lambda! = (\boldsymbol{\lambda}_1!)^{r_1} (\boldsymbol{\lambda}_2!)^{r_2} \cdots\)
\end{itemize}

\hskip-0.5cm \textbf{\emph{Example 1:}} The partitions of
\(\boldsymbol{i}=(2,1)\) are the matrices \[
\begin{pmatrix}
2 \\
1
\end{pmatrix}, \begin{pmatrix}
0 & 2 \\
1 & 0
\end{pmatrix}, \begin{pmatrix}
1 & 1 \\
0 & 1
\end{pmatrix}, \begin{pmatrix}
0 & 1 & 1 \\
1 & 0 & 0
\end{pmatrix} = (\boldsymbol{\lambda}_1, \boldsymbol{\lambda}_2^2),\]
with \[
\boldsymbol{\lambda}_1 = \begin{pmatrix}
0 \\
1
\end{pmatrix} \quad \hbox{and} \quad \boldsymbol{\lambda}_2 = \begin{pmatrix}
1 \\
0
\end{pmatrix}.
\]

The algorithm to get all the partitions of a multi-index resorts to
multiset subdivisions. Let's start by recalling the notion of multiset.
A multiset \(M\) is a ``set with multiplicities''. Suppose \(a \in M.\)
Then the multiplicity of \(a\) is the number of times \(a\) occurs in
\(M\) as a member. For example, the integers \(3\) and \(2\) are the
multiplicities of \(a\) and \(b\) respectively in \(M = \{a,a,a,b,b\}.\)
A subdivision of the multiset \(M\) is a multiset of sub-multisets of
\(M,\) such that their disjoint union returns \(M\). Examples of
subdivisions of \(M = \{a,a,a,b,b\}\) are \begin{equation}
S_1 = \{\{a\},\{a,b\},\{a,b\}\}, \qquad 
S_2 = \{\{a\},\{a,a,b\},\{b\}\}, 
\label{subdivision}
\end{equation} \begin{equation}
S_3  = \{\{a\},\{a,a\},\{b\},\{b\}\}.
\label{subdivision1}
\end{equation} The subdivisions of the multiset \(M = \{a,a,a,b,b\}\)
are in one-to-one correspondence with the partitions
\(\Lambda \vdash (3,2).\) For example, the subdivisions
\(\eqref{subdivision}\) correspond to the partitions
\(\Lambda_1 = (\boldsymbol{\lambda}_2, \boldsymbol{\lambda}_3^2)\) and
\(\Lambda_2 = (\boldsymbol{\lambda}_1, \boldsymbol{\lambda}_2,\boldsymbol{\lambda}_5)\)
respectively, while \(\eqref{subdivision1}\) to
\(\Lambda_3 = (\boldsymbol{\lambda}_1^2, \boldsymbol{\lambda}_2,\boldsymbol{\lambda}_4)\)
with
\[ \boldsymbol{\lambda}_1={0 \choose 1} \! \rightarrow \! \{b\} \,\,\, \boldsymbol{\lambda}_2={1 \choose 0}  \! \rightarrow \! \{a\}\]
\[\boldsymbol{\lambda}_3={1 \choose 1}  \! \rightarrow \! \{a,b\} \,\,\, \boldsymbol{\lambda}_4={2 \choose 0}  \! \rightarrow \! \{a,a\} \,\,\,  \boldsymbol{\lambda}_5={2 \choose 1}  \! \rightarrow \!\{a,a,b\}.\]
Multiset subdivisions can be recovered by using collapsing set
partitions \citep{MR2200529}. If the members \(1, 2, 3\) of the set
\(\{ 1, 2, 3, 4, 5\}\) are made indistinguishable from each other and
called \(a\), and \(4\) and \(5\) are made indistinguishable from each
other and called \(b\), then the set \(\{ 1, 2, 3, 4, 5\}\) has
``collapsed'' to the multiset \(M = \{a,a,a,b,b\}.\) Therefore the
subdivisions of \(M\) can be recovered using the same substitution in
the partitions of \(\{ 1, 2, 3, 4, 5\}.\) For example, \(S_1\) in
\eqref{subdivision} can be recovered from \(\{\{1,4\},\{2,5\}, \{3\}\}\)
or \(\{\{3,5\},\{2,4\}, \{1\}\}\) and so on. As this last example shows,
a subdivision might correspond to several partitions. The number of
partitions corresponding to the same subdivision can be computed using
the
\href{https://www.rdocumentation.org/packages/kStatistics/versions/2.0/topics/countP}{\tt countP}
function of the package. However, to find multi-index partitions using
set partitions is not a particularly efficient algorithm since the
computational cost is proportional to the \(n\)-th Bell number, if \(n\)
is the sum of the multi-index components \citep{MR1937238}. The
\href{https://www.rdocumentation.org/packages/kStatistics/versions/2.0/topics/mkmSet}{\tt mkmSet}
function is based on a different strategy which takes into account the
partitions of the multi-index components. In the following, we describe
the main steps by working on an example.

Suppose we want to generate all the partitions of \((3,2).\) Consider
the partitions of \((3,0)\) corresponding to the partitions
\((3), (1,2), (1^3)\) of the integer \(3,\) and the partitions of
\((0,2)\) corresponding to the partitions \((2),(1^2)\) of the integer
\(2,\) that is \begin{equation}
\Lambda_1= \begin{pmatrix}
3 \\
0
\end{pmatrix}\!\!, \, \Lambda_2= \begin{pmatrix}
1 & 2 \\
0 & 0
\end{pmatrix}\!\!, \, \Lambda_3=\begin{pmatrix}
1 & 1 & 1 \\
0 & 0 & 0
\end{pmatrix} \vdash {3 \choose 0}
\label{(firstsubdivisions)}
\end{equation} \begin{equation}
\Lambda_4= \begin{pmatrix}
0 \\
2
\end{pmatrix}\!\!, \, \Lambda_5= \begin{pmatrix}
0 & 0 \\
1 & 1
\end{pmatrix} \vdash {0 \choose 2}.
\label{(firstsubdivisions1)}
\end{equation}

\noindent The following iterated adding-appending rule is thus
implemented.

\hskip-0.5cm 1. Consider the partition \(\Lambda_5\) in
\eqref{(firstsubdivisions1)}.

\hskip-0.5cm 1.1 Add the first column of \(\Lambda_5\) to each column of
\(\Lambda_1, \Lambda_2\) and \(\Lambda_3\) in
\(\eqref{(firstsubdivisions)}\) one by one with the following rules: the
sum must be done only once (if the column has multiplicities greater
than one) taking as reference the first column; the sum can be done only
to columns whose second component is zero and without subsequent
elements (in the same row) greater than or equal to the integer we are
adding. Then we have \begin{equation}
\!\!\!\!\!\!\!\!\!\!\!{\small \Lambda^{(1,1)}_1= \begin{pmatrix}
3 \\
1
\end{pmatrix} \Lambda^{(1,1)}_2 = \begin{pmatrix}
1 & 2 \\
1 & 0
\end{pmatrix} \Lambda^{(2,1)}_2 = \begin{pmatrix}
1 & 2 \\
0 & 1
\end{pmatrix}  \Lambda^{(1,1)}_3=\begin{pmatrix}
1 & 1 & 1 \\
1 & 0 & 0
\end{pmatrix}}. 
\label{steps1}
\end{equation} 1.2 Append the same column to each partition
\(\Lambda_1, \Lambda_2\) and \(\Lambda_3\) in
\(\eqref{(firstsubdivisions1)}\) that is\\
\begin{equation}
{\small \Lambda^{(1,2)}_1= \begin{pmatrix}
3 & 0 \\
0 & 1
\end{pmatrix}  \Lambda^{(1,2)}_2= \begin{pmatrix}
1 & 2 & 0\\
0 & 0 & 1
\end{pmatrix}  \Lambda^{(1,2)}_3=\begin{pmatrix}
1 & 1 & 1 & 0\\
0 & 0 & 0 & 1
\end{pmatrix}}.
\label{steps2}
\end{equation} 1.3 Repeat steps 1.1 and 1.2 for the second column of
\(\Lambda_5\) with respect to the partitions generated in
\(\eqref{steps1}\) and \(\eqref{steps2}:\)\\
\[{\small \begin{array}{lll}
\Lambda^{(1,1)}_1 = \begin{pmatrix}
3 \\
1
\end{pmatrix} & \!\!\!\!\hbox{add} \Rightarrow \hbox{rule out} & \!\!\!\!\!\hbox{append} \Rightarrow \begin{pmatrix}
3 & 0 \\
1 & 1
\end{pmatrix} \\
\Lambda^{(1,1)}_2 = \begin{pmatrix}
1 & 2 \\
1 & 0
\end{pmatrix} &\!\!\!\!\hbox{add} \Rightarrow \begin{pmatrix}
1 & 2 \\
1 & 1
\end{pmatrix} & \!\!\!\!\!\hbox{append} \Rightarrow \begin{pmatrix}
1 & 2 & 0\\
1 & 0 & 1
\end{pmatrix}\\
\Lambda^{(2,1)}_2=\begin{pmatrix}
1 & 2 \\
0 & 1
\end{pmatrix} & \!\!\!\!\hbox{add} \Rightarrow \hbox{rule out} &  \!\!\!\!\!\hbox{append} \Rightarrow \begin{pmatrix}
1 & 2 & 0\\
0 & 1 & 1
\end{pmatrix} \\
\Lambda^{(1,1)}_3 = \begin{pmatrix}
1 & 1 & 1 \\
1 & 0 & 0
\end{pmatrix} &  \!\!\!\!\hbox{add} \Rightarrow \begin{pmatrix}
1 & 1 & 1 \\
1 & 1 & 0
\end{pmatrix} &  \!\!\!\!\!\hbox{append} \Rightarrow \begin{pmatrix}
1 & 1 & 1 & 0 \\
1 & 0 & 0 & 1
\end{pmatrix} \\
\Lambda^{(1,2)}_1 = \begin{pmatrix}
3 & 0 \\
0 & 1
\end{pmatrix} & \!\!\!\!\hbox{add} \Rightarrow \hbox{rule out} &  \!\!\!\!\!\hbox{append} \Rightarrow \begin{pmatrix}
3 & 0 & 0 \\
0 & 1 & 1
\end{pmatrix} \\
\Lambda^{(1,2)}_2 = \begin{pmatrix}
1 & 2 & 0\\
0 & 0 & 1
\end{pmatrix} & \!\!\!\!\hbox{add} \Rightarrow \hbox{rule out} &  \!\!\!\!\!\hbox{append} \Rightarrow \begin{pmatrix}
1 & 2 & 0 & 0\\
0 & 0 & 1 & 1
\end{pmatrix} \\
\Lambda^{(1,2)}_3 = \begin{pmatrix}
1 & 1 & 1 & 0\\
0 & 0 & 0 & 1
\end{pmatrix} & \!\!\!\!\hbox{add} \Rightarrow \hbox{rule out} & \!\!\!\!\! \hbox{append} \Rightarrow \begin{pmatrix}
1 & 1 & 1 & 0 & 0\\
0 & 0 & 0 & 1 & 1
\end{pmatrix}
\end{array}}\] 2. Repeat step 1 for \(\Lambda_4\) in
\(\eqref{(firstsubdivisions1)}:\)\\
\[{\small \begin{array}{lll}
\Lambda_1= \begin{pmatrix}
3 \\
0
\end{pmatrix} & \!\!\!\!\hbox{add} \Rightarrow \begin{pmatrix}
3 \\
2
\end{pmatrix}  & \!\!\!\!\!\hbox{append} \Rightarrow \begin{pmatrix}
3 & 0 \\
0 & 2
\end{pmatrix}\\
\Lambda_2= \begin{pmatrix}
1 & 2 \\
0 & 0
\end{pmatrix} &\!\!\!\!\hbox{add} \Rightarrow \begin{pmatrix}
1 & 2 \\
2 & 0
\end{pmatrix}, \begin{pmatrix}
1 & 2 \\
0 & 2
\end{pmatrix} & \!\!\!\!\!\hbox{append} \Rightarrow \begin{pmatrix}
1 & 2 & 0\\
0 & 0 & 2
\end{pmatrix} \\
\Lambda_3=\begin{pmatrix}
1 & 1 & 1 \\
0 & 0 & 0
\end{pmatrix} & \!\!\!\!\hbox{add} \Rightarrow \begin{pmatrix}
1 & 1 & 1 \\
2 & 0 & 0
\end{pmatrix} &  \!\!\!\!\!\hbox{append} \Rightarrow \begin{pmatrix}
1 & 1 & 1 & 0 \\
0 & 0 & 0 & 2
\end{pmatrix} 
\end{array}}\] More generally, the
\href{https://www.rdocumentation.org/packages/kStatistics/versions/2.0/topics/mkmSet}{\tt mkmSet}
function lists all the partitions \(\Lambda \vdash \boldsymbol{i},\)
with the columns reordered in increasing lexicographic order, together
with the number of set partitions corresponding to the same multi-index
partition, that is \(\boldsymbol{i}!/\Lambda! \mathfrak{m}(\Lambda)!.\)
In the latest version of the
\href{https://cran.r-project.org/web/packages/kStatistics/index.html}{\tt kStatistics}
package, among the input parameters of the
\href{https://www.rdocumentation.org/packages/kStatistics/versions/2.0/topics/mkmSet}{\tt mkmSet}
function, an input flag parameter has been inserted aiming to print the
multi-index partitions in a more compact form. See the following
example.
\\  \\ \noindent 
\textbf{\emph{Example 2:}} To get all the partitions of
\((2,1)\) run

\begin{verbatim}
>mkmSet(c(2,1),TRUE)
[( 0 1 )( 1 0 )( 1 0 ),  1 ]
[( 0 1 )( 2 0 ),  1 ]
[( 1 0 )( 1 1 ),  2 ]   
[( 2 1 ),  1 ] 
\end{verbatim}

\hskip-0.5cm Note that the integers \(1,1,2,1\) correspond to the
coefficients \(2! 1!/\Lambda! \mathfrak{m}(\Lambda)!.\)

When \(m=1,\) the
\href{https://www.rdocumentation.org/packages/kStatistics/versions/2.0/topics/mkmSet}{\tt mkmSet}
function lists all the partitions \(\lambda\) of the integer \(i.\)
Recall that a partition of an integer \(i\) is a sequence
\(\lambda = (\lambda_1, \lambda_2, \ldots)\) of weakly decreasing
positive integers, named parts of \(\lambda,\) such that
\(\lambda_1 + \lambda_2 + \cdots = i.\) A different notation is
\(\lambda = (1^{r_1}, 2^{r_2}, \ldots),\) where \(r_1, r_2, \ldots\) are
the number of parts of \(\lambda\) equal to \(1,2,\ldots\) respectively.
The length of the partition is \(l(\lambda)=r_1 + r_2 + \cdots.\) We
write \(\lambda \vdash i\) to denote that \(\lambda\) is a partition of
\(i.\)
\\  \\ \noindent 
\textbf{\emph{Example 3:}} To get all the partitions of the
integer \(3\) run

\begin{verbatim}
>mkmSet(c(3),TRUE)
[( 1 )( 1 )( 1 ),  1 ]
[( 1 )( 2 ),  3 ]
[( 3 ),  1 ]
\end{verbatim}

The
\href{https://www.rdocumentation.org/packages/kStatistics/versions/2.0/topics/mkmSet}{\tt mkmSet}
function is called by the
\href{https://www.rdocumentation.org/packages/kStatistics/versions/2.0/topics/intPart}{intPart}
function, specifically designed with the purpose of listing only all the
partitions of a given integer in increasing order. The input flag
parameter allows us to print the partitions in a more compact form.
\\  \\ \noindent 
\textbf{\emph{Example 4:}} To get all the partitions of the
integer \(4\) run

\begin{verbatim}
>intPart(4,TRUE)
[ 1 1 1 1 ]
[ 1 1 2 ]
[ 2 2 ]
[ 1 3 ]
[ 4 ]
\end{verbatim}

The
\href{https://www.rdocumentation.org/packages/partitions/versions/1.1-2/topics/parts}{\tt parts}
function of the
\href{https://cran.r-project.org/web/packages/partitions/index.html}{\tt partitions}
package \citep{partitions} lists all the partitions of a given integer,
but in decreasing order. Instead the
\href{https://www.rdocumentation.org/packages/nilde/versions/1.1-6/topics/get.partitions}{\tt get.partitions}
function of the
\href{https://cran.r-project.org/web/packages/nilde/index.html}{\tt nilde}
package \citep{nilde} finds all the partitions of a given integer with a
fixed length \(l(\lambda)\) \citep{Voinov1439641}. If \(l(\lambda)\) is
equal to the given integer, the
\href{https://www.rdocumentation.org/packages/nilde/versions/1.1-6/topics/get.partitions}{\tt get.partitions}
function lists all the partitions in increasing order.

\hypertarget{section}{%
\section{kStatistics}}

The \(i\)-th \(k\)-statistic \(\kappa_i\) is the (unique) symmetric
estimator whose expectation is the \(i\)-th cumulant \(k_i(Y)\) of a
population character \(Y\) and whose variance is a minimum relative to
all other unbiased estimators.

The
\href{https://www.rdocumentation.org/packages/kStatistics/versions/2.0/topics/nKS}{\tt nKS}
function generates the numerical value of the \(i\)-th \(k\)-statistic
starting from a data sample. The computation relies on the following
polynomials \begin{equation}
{\mathcal P}_t(y) = \sum_{j=1}^t y^j S(t,j) (-1)^{j-1} (j-1)! \qquad \hbox{for} \quad t=1, \ldots, i
\label{(ptpol)}
\end{equation} where \(\{S(t,j)\}\) are the Stirling numbers of second
kind, generated trough the
\href{https://www.rdocumentation.org/packages/kStatistics/versions/2.0/topics/nStirling2}{\tt nStirling2}
function. In detail, suppose to have a sample \(\{a_1, \ldots, a_N\}\)
of \(N\) numerical data and denote with \(p_t\) the \(t\)-th power sum
in the numerical data \begin{equation}
p_t(a_1, \ldots, a_N)  = \sum_{j=1}^N a_j^t, \qquad \hbox{for} \,\,  t\geq 1.
\label{ps}
\end{equation} To carry out the numerical value of the \(i\)-th
\(k\)-statistic for \(i \leq N,\) the
\href{https://www.rdocumentation.org/packages/kStatistics/versions/2.0/topics/nKS}{\tt nKS}
function computes the explicit expression of the polynomial of degree
\(i\) \begin{equation}
Q_i(y) = \sum_{\lambda \vdash i} d_{\lambda} {\mathcal P}_{\lambda}(y) p_{\lambda}
\label{(Qi)}
\end{equation} where the sum is over all the partitions
\(\lambda=(1^{r_1},2^{r_2},\ldots) \vdash i,\) and \begin{equation}
\hskip-1.5cm{\small d_{\lambda}=\frac{i!}{(1!)^{r_1} r_1! (2!)^{r_2} r_2! \cdots} \,\,\, {\mathcal P}_{\lambda}(y) =  [{\mathcal P}_{1}(y)]^{r_1} [{\mathcal P}_{2}(y)]^{r_2} \cdots \,\,\, {p}_{\lambda} =  [p_{1}]^{r_1} [p_{2}]^{r_2} \cdots} \!\!\!
\label{dlambda}
\end{equation} with \(\{{\mathcal P}_{t}(y)\}\) and \(\{p_{t}\}\) given
in \(\eqref{(ptpol)}\) and \(\eqref{ps}\) respectively. The final step
is to replace the powers \(y^t\) in the explicit form of the polynomial
\(\eqref{(Qi)}\) with \((-1)^{t-1} (t-1)!/(N)_t\) for \(t=1, \ldots,i.\)
\\  \\ \noindent 
\textbf{\emph{Example 5:}} Using \(\eqref{(Qi)}\) for
\(i=1,\) we have \(Q_1(y) = {\mathcal P}_1(y) p_1 = y \sum_{j=1}^N a_j\)
and plugging \(1/N\) in place of \(y,\) the sample mean is recovered.
Using \(\eqref{(Qi)}\) for \(i=2,\) we have
\[Q_2(y) = {\mathcal P}_2(y) p_2 +  \big({\mathcal P}_1(y) p_1\big)^2 =  y \sum_{j=1}^N a_j^2 + y^2 \bigg( \big(\sum_{j=1}^N a_j\big)^2 - \sum_{j=1}^N a_j^2\bigg)\]
and plugging \(1/N\) in place of \(y\) and \(-1/N(N-1)\) in place of
\(y^2,\) the sample variance is recovered. Compare the values of the
sample mean, computed with the
\href{https://www.rdocumentation.org/packages/kStatistics/versions/2.0/topics/nKS}{\tt nKS}
function and the
\href{https://www.rdocumentation.org/packages/base/versions/3.6.2/topics/mean}{\tt mean}
function, and the sample variance, computed with the
\href{https://www.rdocumentation.org/packages/kStatistics/versions/2.0/topics/nKS}{\tt nKS}
function and the
\href{https://www.rdocumentation.org/packages/stats/versions/3.6.2/topics/cor}{\tt var}
function, for the following dataset:

\begin{verbatim}
> data<-c(16.34, 10.76, 11.84, 13.55, 15.85, 18.20, 7.51, 10.22, 12.52, 14.68, 
16.08, 19.43, 8.12, 11.20, 12.95, 14.77, 16.83, 19.80, 8.55, 11.58, 12.10, 15.02, 
16.83, 16.98, 19.92, 9.47, 11.68, 13.41, 15.35, 19.11)
> nKS(1,data)
[1] 14.02167
> mean(data)
[1] 14.02167
> nKS(2,data)
[1] 12.65007
> var(data)
[1] 12.65007
\end{verbatim}

Using the
\href{https://www.rdocumentation.org/packages/kStatistics/versions/2.0/topics/nKS}{\tt nKS}
function, for instance, the sample skewness and the sample kurtosis can
be computed. Let us recall that the sample skewness is a measure of the
central tendency of a univariate sample and can be computed as
\(\kappa_3/\kappa_2^{3/2}\) where \(\kappa_2\) and \(\kappa_3\) are the
second and the third \(k\)-statistics respectively
\citep{joanes1998comparing}. The sample kurtosis is a measure of the
tail-heaviness of a sample distribution. The sample excess kurtosis is
defined as the sample kurtosis minus 3 and can be computed as
\(\kappa_4/\kappa_2^{2}\) where \(\kappa_2\) and \(\kappa_4\) are the
second and the fourth \(k\)-statistics respectively
\citep{joanes1998comparing}.

\begin{verbatim}
> nKS(3,data)/sqrt(nKS(2,data))^(3/2)
[1] -0.03216229
> nKS(4,data)/nKS(2,data)^2 + 3
[1] 2.114708
\end{verbatim}

\hskip-0.5cm The main steps of the
\href{https://www.rdocumentation.org/packages/kStatistics/versions/2.0/topics/nKS}{\tt nKS}
function are summarized in the following.

\noindent

\rule{13cm}{0.4pt}

\textbf{\emph{Function
\href{https://www.rdocumentation.org/packages/kStatistics/versions/2.0/topics/nKS}{\tt nKS}}}

\begin{quote}
\emph{i)} Compute the power sums \(p_t\) in \eqref{ps} for
\(t=1, \ldots,i.\)
\end{quote}

\begin{quote}
\emph{ii)} Compute \(S(t,j) (-1)^{j-1} (j-1)!\) in \eqref{(ptpol)} for
\(j=1, \ldots,t\) and \(t=1,\ldots,i.\)
\end{quote}

\begin{quote}
\emph{iii)} Using the
\href{https://www.rdocumentation.org/packages/kStatistics/versions/2.0/topics/mkmSet}{\tt mkmSet}
function, compute all the partitions \(\lambda \vdash i.\)
\end{quote}

\begin{quote}
\emph{iv)} For a given partition \(\lambda\), expand the product
\({\mathcal P}_{\lambda}(y)\) in \(\eqref{(Qi)}\) and compute the
coefficient \(d_{\lambda} p_{\lambda}\) of each monomial in \(Q_i(y)\)
using \(\eqref{dlambda}.\)
\end{quote}

\begin{quote}
\emph{v)} For \(t=1, \ldots, i\) multiply \((-1)^{t-1} (t-1)!/(N)_t\)
with the coefficients of the monomial of degree \(t\) carried out at the
previous step and do the sum over all the resulting numerical values.
\end{quote}

\begin{quote}
\emph{vi)} Repeat steps \emph{iv)} and \emph{v)} for all the partitions
\(\lambda\) carried out at step \emph{iii)} and do the sum over all the
resulting numerical values.
\end{quote}

\noindent

\rule{13cm}{0.4pt}

A similar strategy is employed to compute multivariate \(k\)-statistics
(the
\href{https://www.rdocumentation.org/packages/kStatistics/versions/2.0/topics/nKM}{\tt nKM}
function) of a sample data matrix whose columns each represent a
population character. To simplify the notation, in the following we deal
with the case of a bivariate data set
\(\{(a_{1,1},a_{2,1}) \ldots, (a_{1,N},a_{2,N})\}\) of \(N\) paired
numerical data. Denote with \(p_{(s,t)}\) the bivariate power sum in the
paired data \begin{equation}
\!\!\!p_{(s,t)}[(a_{1,1},a_{2,1}), \ldots, (a_{1,N},a_{2,N})] = \sum_{j=1}^N a^s_{1,j} a^t_{2,j}
\quad \hbox{for} \,\, s,t \geq 1.
\label{doubleps}
\end{equation} Suppose \(\boldsymbol{i}=(i_1, i_2)\) with
\(i_1, i_2 \leq N\) and set \(i=i_1+i_2.\) To carry out the numerical
value of the \(\boldsymbol{i}\)-th multivariate \(k\)-statistic, the
\href{https://www.rdocumentation.org/packages/kStatistics/versions/2.0/topics/nKM}{\tt nKM}
function finds the explicit expression of the polynomial
\begin{equation}
{\mathcal Q}_{\boldsymbol{i}}(y) =  \sum_{\Lambda \vdash \boldsymbol{i}} d_{\Lambda}  P_{\Lambda}(y)
p_{\Lambda}
\label{Qi1}
\end{equation} where the sum is over all the partitions
\(\Lambda=(\boldsymbol{\lambda}_1^{r_1}, \boldsymbol{\lambda}_2^{r_2},\ldots) \vdash \boldsymbol{i},\)
and \begin{equation}
\hskip-1.5cm{\small d_{\Lambda} = \frac{\boldsymbol{i}!}{\Lambda! \, \mathfrak{m}(\Lambda)!} \,\, 
P_{\Lambda}(y)  = 
  [{\mathcal P}_{|\boldsymbol{\lambda}_1|}(y)]^{r_1}  [{\mathcal P}_{|\boldsymbol{\lambda}_2|}(y)]^{r_2} \cdots \,\, p_{\Lambda} = 
[p_{\boldsymbol{\lambda}_1}]^{r_1} [p_{\boldsymbol{\lambda}_2}]^{r_2}  \cdots
\label{dlm}}\!\!\!
\end{equation} with \(\{{\mathcal P}_{t}(y)\}\) and \(\{p_{(s,t)}\}\)
given in \(\eqref{(ptpol)}\) and \(\eqref{doubleps}\) respectively. As
for the univariate \(k\)-statistics, the final step consists in
replacing the powers \(y^j\) in the explicit expression of the
polynomial \(\eqref{Qi1}\) with the numerical values
\((-1)^{j-1} (j-1)!/(N)_j\) for \(j=1, \ldots,i.\)
\\  \\ \noindent 
\textbf{\emph{Example 6:}} To estimate the joint cumulant
\(c_{2,1}\) on the following dataset, run

\begin{verbatim}
> data1<-list(c(5.31,11.16),c(3.26,3.26),c(2.35,2.35),c(8.32,14.34),
c(13.48,49.45),c(6.25,15.05),c(7.01,7.01),c(8.52,8.52),c(0.45,0.45),
c(12.08,12.08),c(19.39,10.42))
> nKM(c(2,1),data1)
[1] -23.7379
\end{verbatim}

If the first column are observations of a population character \(X\) and
the second column observations of a population character \(Y,\) then
\(c_{2,1}\) measures how far from connectedness (as opposite to
independence) are \(X^2\) and \(Y\) \citep{MR4129101}. A similar meaning
has the estimation of the joint cumulant \(c_{2,2,2}\) on the following
dataset:

\begin{verbatim}
> data2<-list(c(5.31,11.16,4.23),c(3.26,3.26,4.10),c(2.35,2.35,2.27),
c(4.31,10.16,6.45),c(3.1,2.3,3.2),c(3.20, 2.31, 7.3))
> nKM(c(2,2,2),data2)
[1] 678.1045
\end{verbatim}

\hskip-0.5cm The main steps of the
\href{https://www.rdocumentation.org/packages/kStatistics/versions/2.0/topics/nKM}{\tt nKM}
function are summarized in the following.

\noindent

\rule{13cm}{0.4pt}

\textbf{\emph{Function
\href{https://www.rdocumentation.org/packages/kStatistics/versions/2.0/topics/nKM}{\tt nKM}}}

\begin{quote}
\emph{i)} Compute the bivariate power sums \(p_{(s,t)}\) in
\(\eqref{doubleps}\) for \(s=1, \ldots,i_1\) and \(t=1, \ldots, i_2.\)
\end{quote}

\begin{quote}
\emph{ii)} For \(i=i_1+i_2,\) compute \(S(t,j) (-1)^{j-1} (j-1)!\) in
\(\eqref{(ptpol)}\) for \(j=1, \ldots,t\) and \(t=1,\ldots,i.\)
\end{quote}

\begin{quote}
\emph{iii)} Using the
\href{https://www.rdocumentation.org/packages/kStatistics/versions/2.0/topics/mkmSet}{\tt mkmSet}
function, compute all the partitions \(\Lambda \vdash \boldsymbol{i}.\)
\end{quote}

\begin{quote}
\emph{iv)} For a given partition \(\Lambda\), expand the product
\(P_{\Lambda}(y)\) in \eqref{Qi1} and compute the coefficient
\(d_{\Lambda} p_{\Lambda}\) of each monomial in
\({\mathcal Q}_{\boldsymbol{i}}(y)\) using \(\eqref{dlm}.\)
\end{quote}

\begin{quote}
\emph{v)} For \(j=1, \ldots, i,\) multiply \((-1)^{j-1} (j-1)!/(N)_j\)
with the coefficient of the monomial of degree \(j\) carried out at the
previous step and do the sum over all the resulting numerical values.
\end{quote}

\begin{quote}
\emph{vi)} Repeat steps \emph{iv)} and \emph{v)} for all the partitions
\(\Lambda\) carried out at step \emph{iii)} and do the sum over all the
resulting numerical values.
\end{quote}

\noindent

\rule{13cm}{0.4pt}

\hypertarget{polykays}{%
\section{Polykays}\label{polykays}}

Similarly to \(k\)-statistics, polykays are symmetric unbiased
estimators of cumulant products. More in detail, when evaluated on a
random sample, the \(\boldsymbol{i}\)-th polykay gives an estimation of
the product \(k_{i_1}(Y) \cdots k_{i_m}(Y),\) where
\(\boldsymbol{i} = (i_1, \ldots, i_m) \in {\mathbb N}_0^m\) and
\(\{k_{i_j}(Y)\}\) are cumulants of a population character \(Y.\)

To simplify the notation, in the following we show how to compute the
\(\boldsymbol{i}\)-th polykay of \(N\) numerical data
\(\{a_1, \ldots, a_N\}\) using the
\href{https://www.rdocumentation.org/packages/kStatistics/versions/2.0/topics/nPS}{\tt nPS}
function for \(\boldsymbol{i} = (i_1, i_2)\). Set \(i=i_1+i_2\) and
suppose \(i \leq N.\) The computation relies on the so-called
logarithmic polynomials \begin{equation}\label{Pj}
\hskip-0.5cm \tilde{P}_{t}(y_1, \ldots, y_i) =  \sum_{\lambda \vdash t}  y_{\lambda} d_{\lambda} 
(-1)^{l(\lambda)-1} (l(\lambda)-1)! 
\end{equation} for \(t=1, \ldots, i\) where the sum is over all the
partitions \(\lambda=(1^{r_1},2^{r_2},\ldots) \vdash t,\)
\(d_{\lambda}\) is given in \(\eqref{dlambda}\) and
\(y_{\lambda} = y_1^{r_1} y_2^{r_2} \cdots.\) To compute the polykay of
order \((i_1, i_2)\), the
\href{https://www.rdocumentation.org/packages/kStatistics/versions/2.0/topics/nPS}{\tt nPS}
function finds the explicit expression of the polynomial
\begin{equation}\label{Qi2}
A_i (y_1, \ldots, y_i) = \sum_{\lambda \vdash i} d_{\lambda} \tilde{P}_{\lambda}(y_1, \ldots, y_i) p_{\lambda} 
\end{equation} where the sum is over all the partitions
\(\lambda=(1^{r_1},2^{r_2},\ldots) \vdash i,\) \(d_{\lambda}\) and
\(p_{\lambda}\) are given in \eqref{dlambda} and
\[\tilde{P}_{\lambda}(y_1, \ldots, y_i)  = [\tilde{P}_{1}(y_1, \ldots, y_i)]^{r_1} [\tilde{P}_{2}(y_1, \ldots, y_i)]^{r_2} \cdots\]
with \(\{\tilde{P}_{t}(y_1, \ldots, y_i)\}\) given in \(\eqref{Pj}.\)
Note that the monomials in \(A_i (y_1, \ldots, y_i)\) are of type
\(y_{\lambda}= y_1^{r_1} y_2^{r_2} \cdots\) with
\(\lambda = (1^{r_1}, 2^{r_2}, \ldots) \vdash i.\) The final step is to
plug suitable numerical values in place of \(y_{\lambda}\) depending on
how the partition \(\lambda\) is constructed. Indeed, set
\begin{equation} 
\hskip-0.5cm{\small \tilde{q}(i_1,i_2) = \bigg\{ \lambda^{\prime} + \lambda^{\prime \prime} \vdash i  \, \big| \, \lambda^{\prime}=(1^{s_1}, 2^{s_2}, \ldots) \vdash i_1, \lambda^{\prime \prime} = (1^{t_1}, 2^{t_2},\ldots) \vdash i_2\bigg\}} \!\!\!
\label{pol2}
\end{equation} where
\(\lambda^{\prime}+\lambda^{\prime \prime} = (1^{r_1}, 2^{r_2}, \ldots)\)
with \(r_j = s_j + t_j\) for \(j=1,2,\ldots.\) Then \(y_{\lambda}\) is
replaced by \(0\) if \(\lambda \not \in \tilde{q}(i_1,i_2)\) otherwise
by \begin{equation} 
\frac{(-1)^{l(\lambda^{\prime})-1}(l(\lambda^{\prime})-1)! (-1)^{l(\lambda^{\prime \prime})-1}(l(\lambda^{\prime \prime})-1)!}{(N)_{l(\lambda^{\prime \prime})+l(\lambda^{\prime \prime})}} \frac{d_{\lambda^{\prime}} d_{\lambda^{\prime \prime}}}{d_{\lambda^{\prime}+\lambda^{\prime \prime}}}.
\label{pol1}
\end{equation} Note that \(d_{\lambda^{\prime}}\) and
\(d_{\lambda^{\prime \prime}}\) in \eqref{pol1} are recovered from
\(\eqref{dlambda}.\)
\\  \\ \noindent 
\textbf{\emph{Example 7:}} Suppose we need to estimate the
square of the variance \(\sigma^2\) of the population character \(Y\)
from which the data of Example 5 have been sampled. We have

\begin{verbatim}
> nKS(2,data)^2
[1] 160.0243
> var(data)^2
[1] 160.0243
\end{verbatim}

\hskip-0.5cm but \(k_2^2\) is not an unbiased estimator of the square of
\(\sigma^2.\) An unbiased estimator of such a square is the polykay of
order \((2,2),\) that is

\begin{verbatim}
> nPS(c(2,2),data)
[1] 154.1177
\end{verbatim}

\hskip-0.5cm The main steps of the
\href{https://www.rdocumentation.org/packages/kStatistics/versions/2.0/topics/nPS}{\tt nPS}
function are summarized in the following.

\noindent

\rule{13cm}{0.4pt}

\textbf{\emph{Function
\href{https://www.rdocumentation.org/packages/kStatistics/versions/2.0/topics/nPS}{\tt nPS}}}

\begin{quote}
\emph{i)} Set \(i=i_1+i_2\) and compute the power sums \(p_t\) in
\eqref{ps} for \(t=1, \ldots,i.\)
\end{quote}

\begin{quote}
\emph{ii)} Generate the polynomials \(\tilde{P}_{t}(y_1, \ldots, y_i)\)
in \(\eqref{Pj}\) for \(t=1, \ldots, i.\)
\end{quote}

\begin{quote}
\emph{iii)} Using the
\href{https://www.rdocumentation.org/packages/kStatistics/versions/2.0/topics/mkmSet}{\tt mkmSet}
function, compute all the partitions \(\lambda \vdash i.\)
\end{quote}

\begin{quote}
\emph{iv)} For a given partition \(\lambda\), expand the product
\(\tilde{P}_{\lambda}(y_1, \ldots, y_i)\) in \(\eqref{Qi2};\) then plug
\(\eqref{pol1}\) or \(0\) in each monomial \(y_{\lambda},\) depending if
\(\lambda\) is or not in the set \(\tilde{q}(i_1,i_2)\) given in
\(\eqref{pol2}.\)
\end{quote}

\begin{quote}
\emph{v)} Multiply the numerical value of \(\tilde{P}_{\lambda}\)
carried out at step \emph{iv)} with \(d_{\lambda} p_{\lambda}\) given in
\eqref{dlambda}.
\end{quote}

\begin{quote}
\emph{vi)} Repeat steps \emph{iv)} and \emph{v)} for all the partitions
\(\lambda\) carried out at step \emph{iii)} and do the sum over all the
resulting numerical values.
\end{quote}

\noindent

\rule{13cm}{0.4pt}

Multivariate polykays are unbiased estimators of products of
multivariate cumulants and the
\href{https://www.rdocumentation.org/packages/kStatistics/versions/2.0/topics/nPM}{\tt nPM}
function returns a numerical value for these estimators when evaluated
on a random sample. As before, to show how the
\href{https://www.rdocumentation.org/packages/kStatistics/versions/2.0/topics/nPM}{\tt nPM}
function works, we consider a bivariate sample of \(N\) numerical data,
that is \(\{(a_{1,1},a_{2,1}) \ldots, (a_{1,N},a_{2,N})\}.\) If we
choose \(\boldsymbol{i}=(i_1, i_2)\) and \(\boldsymbol{j}=(j_1, j_2)\)
with \(i_1 + i_2 + j_1 + j_2 \leq N\) as input of the
\href{https://www.rdocumentation.org/packages/kStatistics/versions/2.0/topics/nPM}{\tt nPM}
function, the output is a numerical value which represents an estimated
value of the product
\(k_{\boldsymbol{i}}(X,Y) k_{\boldsymbol{j}}(X,Y),\) where
\(k_{\boldsymbol{i}}(X,Y)\) and \(k_{\boldsymbol{j}}(X,Y)\) are
cumulants of the population characters \((X,Y).\) The computation relies
on suitable polynomials in the indeterminates \(\{y_{(s,t)}\}\) for
\(s=0,\ldots, w_1, t=0,\ldots, w_2,\) with \(s+t>0\) and
\(w_1=i_1+j_1, w_2=i_2+j_2.\) These polynomials are a multivariable
generalization of \(\eqref{Pj},\) that is \begin{equation}\label{Pjm}
\tilde{P}_{\boldsymbol{k}}\left( \{y_{(s,t)}\} \right)  =  \sum_{\Lambda \vdash \boldsymbol{k}} 
y_{\Lambda} d_{\Lambda} (-1)^{l(\Lambda)-1} (l(\Lambda)-1)!
\end{equation} for
\(\boldsymbol{0} < \boldsymbol{k} \leq \boldsymbol{w}=(w_1,w_2),\) where
the sum is over all the partitions
\(\Lambda=(\boldsymbol{\lambda}_1^{r_1}, \boldsymbol{\lambda}_2^{r_2},\ldots) \vdash \boldsymbol{k}\)
and
\(y_{\Lambda} = y_{\boldsymbol{\lambda}_1}^{r_1} y_{\boldsymbol{\lambda_2}}^{r_2} \cdots.\)
To compute the multivariate polykay of order
\((\boldsymbol{i}, \boldsymbol{j}),\) the
\href{https://www.rdocumentation.org/packages/kStatistics/versions/2.0/topics/nPM}{\tt nPM}
function finds the explicit expression of the polynomial
\begin{equation}\label{Ai}
{\mathcal A}_{\boldsymbol{w}} \left( \{y_{(s,t)}\} \right) =  \sum_{\Lambda \vdash \boldsymbol{w}} d_{\Lambda}  \tilde{P}_{\Lambda}\left( \{y_{(s,t)}\} \right) p_{\Lambda} 
\end{equation} where the sum is over all the partitions
\(\Lambda=(\boldsymbol{\lambda}_1^{r_1}, \boldsymbol{\lambda}_2^{r_2},\ldots) \vdash \boldsymbol{w},\)
\(d_{\Lambda}\) and \(p_{\Lambda}\) are given in \(\eqref{dlm}\), and
\[\tilde{P}_{\Lambda}\left( \{y_{(s,t)}\} \right)  = [\tilde{P}_{\boldsymbol{\lambda}_1}
\left( \{y_{(s,t)}\} \right) ]^{r_1} [\tilde{P}_{\boldsymbol{\lambda}_2}\left( \{y_{(s,t)}\} \right) ]^{r_2} 
\cdots\] with
\(\{\tilde{P}_{\boldsymbol{\lambda}}\left( \{y_{(s,t)}\} \right)\}\)
given in \(\eqref{Pjm}.\) The monomials in
\({\mathcal A}_{\boldsymbol{w}}\left( \{y_{(s,t)}\} \right)\) are of
type \(y_{\Lambda}\) with \(\Lambda \vdash \boldsymbol{w}.\) The final
step is to plug suitable numerical values in place of \(y_{\Lambda}\)
depending on how the partition \(\Lambda\) is constructed. Indeed, set
\begin{equation}\label{qw}
\hskip-1.3cm{\small \tilde{q}(\boldsymbol{w}) = \bigg\{  \Lambda^{\prime}+\Lambda^{\prime \prime} \vdash \boldsymbol{w} \, \big| \, \Lambda^{\prime}=(\boldsymbol{\lambda}_1^{\prime s_1}, \boldsymbol{\lambda}_2^{\prime s_2}, \ldots) \vdash \boldsymbol{i}, \Lambda^{\prime \prime} = (\boldsymbol{\lambda}_1^{\prime \prime t_1}, \boldsymbol{\lambda}_2^{\prime \prime t_2},\ldots) \vdash \boldsymbol{j} \bigg\}},\!\!\!
\end{equation} where
\(\Lambda^{\prime}+\Lambda^{\prime \prime}=(\tilde{\boldsymbol{\lambda}}_1^{r_1}, \tilde{\boldsymbol{\lambda}}_2^{r_2}, \ldots)\)
is built with the columns of \(\Lambda^{\prime}\) and
\(\Lambda^{\prime \prime}\) rearranged in increasing lexicographic order
and such that \(r_j=s_j\) if
\(\tilde{\boldsymbol{\lambda}}_j = \boldsymbol{\lambda}^{\prime}_j\) or
\(r_j=t_j\) if
\(\tilde{\boldsymbol{\lambda}}_j = \boldsymbol{\lambda}^{\prime \prime}_j\)
or \(r_j=s_j+t_j\) if
\(\tilde{\boldsymbol{\lambda}}_j = \boldsymbol{\lambda}^{\prime}_j = \boldsymbol{\lambda}^{\prime \prime}_j.\)
Therefore in the explicit expression of \(\eqref{Ai},\) \(y_{\Lambda}\)
is replaced by \(0\) if \(\Lambda \not \in \tilde{q}(\boldsymbol{w})\)
otherwise by \begin{equation}\label{ybmtau}
\frac{{(-1)^{l(\Lambda^{\prime})-1}(l(\Lambda^{\prime})-1)!(-1)^{l(\Lambda^{\prime \prime})-1}(l(\Lambda^{\prime \prime})-1)!}}{(N)_{l(\Lambda^{\prime})+l(\Lambda^{\prime \prime})}} \frac{d_{\Lambda^{\prime}} d_{\Lambda^{\prime \prime}}}{d_{\Lambda^{\prime}+\Lambda^{\prime \prime}}}.
\end{equation} Note that \(d_{\Lambda^{\prime}}\) and
\(d_{\Lambda^{\prime \prime}}\) in \(\eqref{ybmtau}\) are recovered from
\(\eqref{dlm}.\)
\\  \\ \noindent 
\textbf{\emph{Example 8:}} For the same dataset employed in
Example 6, to estimate \(k_{(2,1)}(X,Y) k_{(1,0)}(X,Y)\) run

\begin{verbatim}
> nPM(list(c(2,1),c(1,0)),data1)
[1] 48.43243
\end{verbatim}

\hskip-0.5cm The main steps of the
\href{https://www.rdocumentation.org/packages/kStatistics/versions/2.0/topics/nPM}{\tt nPM}
function are summarized in the following.

\noindent

\rule{13cm}{0.4pt}

\textbf{\emph{Function
\href{https://www.rdocumentation.org/packages/kStatistics/versions/2.0/topics/nPM}{nPM}}}

\begin{quote}
\emph{i)} Set \(w_1=i_1+j_1\) and \(w_2=i_2+j_2;\) compute the power
sums \(p_{(s,t)}\) in \eqref{doubleps} for \(s=1, \ldots, w_1\) and
\(t=1, \ldots, w_2.\)
\end{quote}

\begin{quote}
\emph{ii)} Generate the polynomials
\(\tilde{P}_{\boldsymbol{k}} \left( \{y_{(s,t)}\} \right)\) in
\(\eqref{Pjm}\) for \(0<\boldsymbol{k} \leq \boldsymbol{w}=(w_1,w_2).\)
\end{quote}

\begin{quote}
\emph{iii)} Using the
\href{https://www.rdocumentation.org/packages/kStatistics/versions/2.0/topics/mkmSet}{\tt mkmSet}
function, compute all the partitions \(\Lambda \vdash \boldsymbol{w}.\)
\end{quote}

\begin{quote}
\emph{iv)} For a given partition \(\Lambda\), expand the product
\(\tilde{P}_{\Lambda}\left( \{y_{(s,t)}\}\right)\) in \(\eqref{Ai}\) and
plug \(\eqref{ybmtau}\) or \(0\) in each obtained monomial of type
\(y_{\Lambda}\) depending if \(\Lambda\) is or not in
\(\tilde{q}(\boldsymbol{w})\) given in \(\eqref{qw}.\)
\end{quote}

\begin{quote}
\emph{v)} Multiply the numerical value of \(\tilde{P}_{\Lambda}\)
obtained at step \emph{iv)} with \(d_{\Lambda} p_{\Lambda}\) given in
\(\eqref{dlm}.\)
\end{quote}

\begin{quote}
\emph{vi)} Repeat steps \emph{iv)} and \emph{v)} for all the partitions
\(\Lambda\) carried out at step \emph{iii)} and do the sum over all the
resulting numerical values.
\end{quote}

\noindent

\rule{13cm}{0.4pt}

\noindent \textbf{\emph{Remark 1:}} The master
\href{https://www.rdocumentation.org/packages/kStatistics/versions/2.0/topics/nPolyk}{\tt nPolyk}
function runs one of the
\href{https://www.rdocumentation.org/packages/kStatistics/versions/2.0/topics/nKS}{\tt nKS},
\href{https://www.rdocumentation.org/packages/kStatistics/versions/2.0/topics/nKM}{\tt nKM},
\href{https://www.rdocumentation.org/packages/kStatistics/versions/2.0/topics/nPS}{\tt nPS}
and
\href{https://www.rdocumentation.org/packages/kStatistics/versions/2.0/topics/nPM}{\tt nPM}
functions depending if we ask for simple \(k\)-statistics, multivariate
\(k\)-statistics, simple polykays or multivariate polykays.

\hypertarget{bell-polynomials-and-generalizations}{%
\section{Bell polynomials and
generalizations}\label{bell-polynomials-and-generalizations}}

The algorithms to produce \(k\)-statistics and polykays rely on handling
suitable polynomial families which are special cases of generalizations
of Bell polynomials, as introduced in this section. Moreover, there are
further families of polynomials widely used in applications which are
special cases of these polynomials. For the most popular ones, we have
implemented special functions in the
\href{https://cran.r-project.org/web/packages/kStatistics/index.html}{\tt kStatistics}
package. The list is not exhaustive, see for instance \citet{MR741185}.
Furthermore additional families of polynomials might be recovered using
the multivariate Faà di Bruno's formula. We will give some examples in
the next section.

The \(\boldsymbol{i}\)-th generalized (complete exponential) Bell
polynomial in the indeterminates \(y_1, \ldots, y_n\) is
\begin{equation}
\hskip-0.5cm{\small h_{\boldsymbol{i}}(y_1, \ldots, y_n) = \boldsymbol{i}! \sum_{{\Lambda} \vdash \boldsymbol{s}_1, \ldots, \tilde{\Lambda} \vdash  \boldsymbol{s}_n \atop \boldsymbol{s}_1 + \cdots + \boldsymbol{s}_n = \boldsymbol{i}} y_1^{l(\Lambda)} \cdots y_n^{l(\tilde{\Lambda})} \frac{g_{1,\Lambda} \cdots g_{n,\tilde{\Lambda} }}{\Lambda!
\cdots \tilde{\Lambda}! \, \mathfrak{m}(\Lambda)! \cdots \mathfrak{m}(\tilde{\Lambda})!}}\!\!\!
\label{(sol22ter)}
\end{equation} where the sum is over all the partitions
\({\Lambda} \vdash \boldsymbol{s}_1, \ldots, \tilde{\Lambda} \vdash \boldsymbol{s}_n\)
with \((\boldsymbol{s}_1,\ldots,\boldsymbol{s}_n)\) \(m\)-tuples of
non-negative integers such that
\(\boldsymbol{s}_1 + \cdots + \boldsymbol{s}_n = \boldsymbol{i}\) and
\begin{equation}
\begin{array}{rcl}
g_{1,\Lambda} & = & g_{1; \boldsymbol{\lambda}_1}^{r_1} g_{1; \boldsymbol{\lambda}_2}^{r_2} \cdots \qquad \hbox{for}  \, \Lambda=(\boldsymbol{\lambda}_1^{r_1} , \boldsymbol{\lambda}_2^{r_2}, \ldots) \\
& \vdots & \\
g_{n,\tilde{\Lambda}} & = &  g_{n; \tilde{\boldsymbol{\lambda}}_1}^{t_1} g_{n; \tilde{\boldsymbol{\lambda}}_2}^{t_2} \cdots \qquad \hbox{for}  \,\, \tilde{\Lambda}=(\tilde{\boldsymbol{\lambda}}_1^{t_1} , \tilde{\boldsymbol{\lambda}}_2^{t_2}, \ldots)
\label{sequences}
\end{array}
\end{equation} with
\(\{ g_{1; \boldsymbol{\lambda}}\}, \ldots, \{ g_{n; \boldsymbol{\lambda}}\}\)
multi-indexed sequences. These polynomials are the output of the
\href{https://www.rdocumentation.org/packages/kStatistics/versions/2.0/topics/GCBellPol}{\tt GCBellPol}
function.
\\  \\ \noindent 
\textbf{\emph{Example 9:}} To get \(h_{(1,1)}(y_1, y_2)\)
run

\begin{verbatim}
> GCBellPol(c(1,1),2)
[1] (y1)(y2)g1[0,1]g2[1,0] + (y1)(y2)g1[1,0]g2[0,1] + (y1^2)g1[0,1]g1[1,0] +
(y1)g1[1,1] + (y2^2)g2[0,1]g2[1,0] + (y2)g2[1,1]
\end{verbatim}

The
\href{https://www.rdocumentation.org/packages/kStatistics/versions/2.0/topics/e_GCBellPol}{\tt e\_GCBellPol}
function evaluates \(h_{\boldsymbol{i}}(y_1, \ldots, y_n)\) when its
indeterminates \(y_1, \ldots, y_n\) and/or its coefficients are
substituted with numerical values.
\\  \\ \noindent 
\textbf{\emph{Example 10:}} To plug the values from \(1\) to
\(6\) respectively into the coefficients \texttt{g1{[}\ ,\ {]}} and
\texttt{g2{[}\ ,\ {]}} of the polynomial \(h_{(1,1)}(y_1, y_2)\) given
in Example 9 run

\begin{verbatim}
> e_GCBellPol(c(1,1), 2, "g1[0,1]=1, g1[1,0]=2, g1[1,1]=3, g2[0,1]=4, g2[1,0]=5, 
g2[1,1]=6")
[1] 13(y1)(y2) + 2(y1^2) + 3(y1) + 20(y2^2) + 6(y2)
\end{verbatim}

\noindent To evaluate \(h_{(1,1)}(1, 5)\) run

\begin{verbatim}
> e_GCBellPol(c(1,1), 2, "y1=1,  y2=5, g1[0,1]=1, g1[1,0]=2, g1[1,1]=3, g2[0,1]=4, 
g2[1,0]=5, g2[1,1]=6")
[1] 600
\end{verbatim}

When the multi-indexed sequences
\(\{ g_{1; \boldsymbol{\lambda}}\}, \ldots, \{ g_{n; \boldsymbol{\lambda}}\}\)
are all equal, the number of distinct addends in \(\eqref{(sol22ter)}\)
might reduce and the corresponding generalized Bell polynomial is
denoted by \(\tilde{h}_{\boldsymbol{i}}(y_1, \ldots, y_n)\). To deal
with this special case, we have inserted an input flag parameter in the
\href{https://www.rdocumentation.org/packages/kStatistics/versions/2.0/topics/e_GCBellPol}{\tt e\_GCBellPol}
function.
\\  \\ \noindent 
\textbf{\emph{Example 11:}} To compare
\(\tilde{h}_{(1,1)}(y_1, y_2)\) with \(h_{(1,1)}(y_1, y_2)\) given in
Example 9 run

\begin{verbatim}
> GCBellPol(c(1,1),2,TRUE)
[1] 2(y1)(y2)g[0,1]g[1,0] + (y1^2)g[0,1]g[1,0] + (y1)g[1,1] + (y2^2)g[0,1]g[1,0] +
(y2)g[1,1]
\end{verbatim}

Set \(n=1\) in \(\eqref{(sol22ter)}\). Then
\(h_{\boldsymbol{i}}(y_1, \ldots, y_n)\) reduces to the univariate
polynomial\\
\begin{equation}
h_{\boldsymbol{i}}(y) =  \sum_{{\Lambda} \vdash  \boldsymbol{i}} y^{l(\Lambda)} d_{\Lambda}  
g_{\Lambda}
\label{(redGC)}
\end{equation} where the sum is over all the partitions
\(\Lambda=(\boldsymbol{\lambda}_1^{r_1} , \boldsymbol{\lambda}_2^{r_2}, \ldots) \vdash \boldsymbol{i},\)
\(d_{\Lambda}\) is given in \(\eqref{dlm}\) and
\(g_{\Lambda} = g_{\boldsymbol{\lambda}_1}^{r_1} g_{\boldsymbol{\lambda}_2}^{r_2} \cdots.\)
\\  \\ \noindent 
\textbf{\emph{Example 12:}} To get \(h_{(1,1)}(y)\) run

\begin{verbatim}
> GCBellPol(c(1,1),1)
[1] (y^2)g[0,1]g[1,0] + (y)g[1,1]
\end{verbatim}
\noindent
\textbf{\emph{Remark 2:}} For all
\(\boldsymbol{i} \in {\mathbb N}_0^m,\) we have
\(h_{\boldsymbol{i}}(y_1 + \cdots + y_n)= \tilde{h}_{\boldsymbol{i}}(y_1, \ldots, y_n),\)
where \(\tilde{h}_{\boldsymbol{i}}(y_1, \ldots, y_n)\) is the
\(\boldsymbol{i}\)-th generalized Bell polynomial \(\eqref{(sol22ter)}\)
corresponding to all equal multi-indexed sequences
\(\{ g_{1, \boldsymbol{\lambda}}\}, \ldots, \{ g_{n, \boldsymbol{\lambda}}\}\)
\citep{MR3437172}. Therefore the
\href{https://www.rdocumentation.org/packages/kStatistics/versions/2.0/topics/e_GCBellPol}{\tt e\_GCBellPol}
function, with the input flag \texttt{TRUE}, produces also an explicit
expression of \(h_{\boldsymbol{i}}(y_1 + \cdots + y_n).\)

The algorithm to generate joint moments in terms of joint cumulants and
viceversa follows the same pattern designed to generate
\(\{h_{\boldsymbol{i}}(y)\}.\) Indeed if
\(\{k_{\boldsymbol{i}}(\boldsymbol{Y})\}\) and
\(\{m_{\boldsymbol{i}}(\boldsymbol{Y})\}\) denote the sequences of joint
cumulants and joint moments of a random vector \(\boldsymbol{Y}\)
respectively, then \begin{equation}\label{cummom}
\hskip-1.3cm{\small m_{\boldsymbol{i}}(\boldsymbol{Y}) = \sum_{{\Lambda} \vdash  \boldsymbol{i}} d_{\Lambda} k_{\Lambda}(\boldsymbol{Y}) \,\,  \hbox{and} \,\, k_{\boldsymbol{i}}(\boldsymbol{Y}) = \sum_{{\Lambda} \vdash  \boldsymbol{i}} (-1)^{l(\Lambda)-1} (l(\Lambda)-1)! d_{\Lambda} m_{\Lambda}(\boldsymbol{Y})},
\end{equation} where the sum is over all the partitions
\(\Lambda=(\boldsymbol{\lambda}_1^{r_1} , \boldsymbol{\lambda}_2^{r_2}, \ldots) \vdash \boldsymbol{i},\)
\(d_{\Lambda}\) is given in \(\eqref{dlm}\) and
\[{\small m_{\Lambda}(\boldsymbol{Y})= [m_{{\boldsymbol{\lambda}}_1}(\boldsymbol{Y})]^{r_1} [m_{{\boldsymbol{\lambda}}_2}(\boldsymbol{Y})]^{r_2} \cdots \quad k_{\Lambda}(\boldsymbol{Y})= [k_{{\boldsymbol{\lambda}}_1}(\boldsymbol{Y})]^{r_1} [k_{{\boldsymbol{\lambda}}_2}(\boldsymbol{Y})]^{r_2} \cdots}.\]
In particular

\begin{itemize}
\item
  the
  \href{https://www.rdocumentation.org/packages/kStatistics/versions/2.0/topics/mom2cum}{\tt mom2cum}
  function returns the right hand side of the first equation in
  \(\eqref{cummom}\), using the same algorithm producing
  \(h_{\boldsymbol{i}}(y)\) in \(\eqref{(redGC)}\) with the sequence
  \(\{k_{\boldsymbol{\lambda}}\}\) in place of
  \(\{g_{\boldsymbol{\lambda}}\}\) and with \(1\) in place of \(y;\)
\item
  the
  \href{https://www.rdocumentation.org/packages/kStatistics/versions/2.0/topics/cum2mom}{\tt cum2mom}
  function returns the right hand side of the latter equation in
  \(\eqref{cummom}\), using the same algorithm producing
  \(h_{\boldsymbol{i}}(y)\) in \eqref{(redGC)} with the sequence
  \(\{m_{\boldsymbol{\lambda}}\}\) in place of
  \(\{g_{\boldsymbol{\lambda}}\}\) and with \((-1)^{j-1} (j-1)!\) in
  place of the powers \(y^j\) for \(j=1,\ldots,|\boldsymbol{i}|.\)
\end{itemize}

When the multi-index \(\boldsymbol{i}\) reduces to an integer \(i,\)
formulae \(\eqref{cummom}\) are the classical expressions of univariate
moments in terms of univariate cumulants and viceversa: the
\href{https://www.rdocumentation.org/packages/kStatistics/versions/2.0/topics/mom2cum}{\tt mom2cum}
and
\href{https://www.rdocumentation.org/packages/kStatistics/versions/2.0/topics/cum2mom}{cum2mom}
functions do the same when the input is an integer.
\\  \\ \noindent 
\textbf{\emph{Example 13:}} To get \(m_{(3,1)}\) in terms of
\(k_{(i,j)}\) run

\begin{verbatim}
> mom2cum(c(3,1))
[1] k[0,1]k[1,0]^3 + 3k[0,1]k[1,0]k[2,0] + k[0,1]k[3,0] + 3k[1,0]^2k[1,1] +
3k[1,0]k[2,1] + 3k[1,1]k[2,0] + k[3,1]
\end{verbatim}

\noindent To get \(k_{(3,1)}\) in terms of \(m_{(i,j)}\) run

\begin{verbatim}
> cum2mom(c(3,1))
[1]  - 6m[0,1]m[1,0]^3 + 6m[0,1]m[1,0]m[2,0] - m[0,1]m[3,0] + 
6m[1,0]^2m[1,1] - 3m[1,0]m[2,1] - 3m[1,1]m[2,0] + m[3,1]
\end{verbatim}
\noindent
\textbf{\emph{Remark 3:}} There are different functions in
\texttt{R} performing similar computations for cumulants and moments:
for instance see \citet{Leeuw} for the multivariate case. A different
strategy would rely on the recursive relations between cumulants and
moments \citep{MR3809541}.

Similarly to \(\eqref{cummom}\), some of the polynomials employed in the
previous sections are generated using the same pattern developed to find
the explicit expression of \(h_{\boldsymbol{i}}(y)\) in
\(\eqref{(redGC)}\):

\begin{itemize}
\item
  the generation of an explicit expression of
  \({\mathcal Q}_{\boldsymbol{i}}(y)\) in \(\eqref{Qi1}\) parallels the
  one implemented for \(h_{\boldsymbol{i}}(y)\) with \(1\) in place of
  \(y\) and with the polynomial sequence
  \(\{{\mathcal P}_{|\boldsymbol{\lambda}|}(y) p_{\boldsymbol{\lambda}}\}\)
  in place of the sequence \(\{g_{\boldsymbol{\lambda}}\};\)
\item
  the same for the polynomials
  \(\tilde{P}_{\boldsymbol{k}}\left( \{y_{(s,t)}\} \right)\) in
  \(\eqref{Pjm}\) with \((-1)^{j-1} (j-1)!\) for
  \(j=1,\ldots,|\boldsymbol{i}|\) in place of the powers \(y^j\) and
  with the polynomial sequence \(\{y_{\boldsymbol{\lambda}}\}\) in place
  of the sequence \(\{g_{\boldsymbol{\lambda}}\};\)
\item
  the same for the polynomials
  \({\mathcal A}_{\boldsymbol{w}} \left( \{y_{(s,t)}\} \right)\) in
  \(\eqref{Ai}\) with \(1\) in place of \(y\) and with the polynomial
  sequence
  \(\{\tilde{P}_{\boldsymbol{\lambda}}\left( \{y_{(s,t)}\} \right)p_{\boldsymbol{\lambda}}\}\)
  in place of the sequence \(\{g_{\boldsymbol{\lambda}}\}.\)
\end{itemize}

Note that when the multi-index \(\boldsymbol{i}\) in \(\eqref{(redGC)}\)
reduces to a positive integer \(i,\) then the polynomial
\(h_{\boldsymbol{i}}(y)\) becomes \begin{equation}
h_i(y) = \sum_{\lambda \vdash i}  d_{\lambda}  y^{l(\lambda)} g_{\lambda} 
\label{(ffaal)}
\end{equation} where the sum is over all the partitions
\(\lambda=(1^{r_1}, 2^{r_2}, \ldots) \vdash i,\) \(d_{\lambda}\) is
given in \(\eqref{dlambda}\) and
\(g_{\lambda}= g_1^{r_1} g_2^{r_2} \ldots\) with \(\{g_j\}\) a suitable
sequence.
\\  \\ \noindent 
\textbf{\emph{Example 14:}} To get \(h_{3}(y)\) run

\begin{verbatim}
> GCBellPol(c(3),1)
[1] (y^3)g[1]^3 + 3(y^2)g[1]g[2] + (y)g[3]
\end{verbatim}

With a combinatorial structure very similar to \(\eqref{(ffaal)}\), the
\(i\)-th general partition polynomial has the following expression in
the indeterminates \(y_1, \ldots, y_i\): \begin{equation}
G_i( a_1, \ldots, a_i; y_1, \ldots, y_i) = \sum_{\lambda \vdash i}  d_{\lambda}  a_{l(\lambda)} y_{\lambda} 
\label{(ffaa)}
\end{equation} where the sum is over all the partitions
\(\lambda=(1^{r_1}, 2^{r_2}, \ldots) \vdash i,\) \(d_{\lambda}\) is
given in \(\eqref{dlambda},\) \(\{a_j\}\) is a suitable numerical
sequence and \(y_{\lambda} = y_1^{r_1} y_2^{r_2} \ldots.\) It's a
straightforward exercise to prove that \begin{equation}
G_i( a_1, \ldots, a_i; y_1, \ldots, y_i) = \sum_{j=1}^i a_j B_{i,j}(y_1, \ldots, y_{i-j+1}), 
\label{gpp}
\end{equation} where \(\{B_{i,j}\}\) are the (partial) exponential Bell
polynomials \begin{equation}
B_{i,j}(y_1,  \ldots, y_{i-j+1}) =  \sum_{\bar{p}(i,j)} d_{\lambda} y_{\lambda} 
\label{(parexpBell)}
\end{equation} where
\(\bar{p}(i,j) = \{\lambda=(1^{r_1}, 2^{r_2}, \ldots) \vdash i | l(\lambda)=j\},\)
\(d_{\lambda}\) is given in \(\eqref{dlambda}\) and
\(y_{\lambda} =y_1^{r_1} y_2^{r_2} \cdots.\) The polynomials in
\(\eqref{(ffaa)}\) are widely used in applications such as
combinatorics, probability theory and statistics \citep{MR1937238}. As
particular cases, they include the exponential polynomials and their
inverses, the logarithmic polynomials \(\eqref{Pj}\), the potential
polynomials and many others \citep{MR741185}.

The general partition polynomials are the output of the
\href{https://www.rdocumentation.org/packages/kStatistics/versions/2.0/topics/gpPart}{\tt gpPart}
function.
\\  \\ \noindent 
\textbf{\emph{Example 15:}} To get
\(G_4( a_1, a_2, a_3, a_4; y_1, y_2, y_3, y_4)\) run

\begin{verbatim}
> gpPart(4)
[1] a4(y1^4) + 6a3(y1^2)(y2) + 3a2(y2^2) + 4a2(y1)(y3) + a1(y4)
\end{verbatim}

When \(a_1 = \ldots = a_i =1,\) the \(i\)-th general partition
polynomial in \(\eqref{gpp}\) reduces to the complete (exponential) Bell
polynomial \begin{equation}
G_i( 1, \ldots, 1; y_1, \ldots, y_i) =  \sum_{j=1}^i B_{i,j}(y_1, \ldots, y_{i-j+1})
\label{BC}
\end{equation} where \(\{B_{i,j}\}\) are the (partial) exponential Bell
polynomials \(\eqref{(parexpBell)}\). For instance, the polynomials
\(Q_i(y)\) in \(\eqref{(Qi)}\) are generated using the same pattern
developed to generate \(\eqref{BC}\) with \({\mathcal P}_{j}(y) p_{j}\)
in place of \(y_j.\)

The
\href{https://www.rdocumentation.org/packages/kStatistics/versions/2.0/topics/eBellPol}{\tt eBellPol}
function returns the complete (exponential) Bell polynomials
\(\eqref{BC}\). The same function also produces the (partial)
exponential Bell polynomial \(B_{i,j}(y_1, \ldots, y_{i-j+1})\) using
\(\eqref{(ffaa)}\) with \(a_k=\delta_{k,j}\) (the Kronecker delta) for
\(k=1, \ldots,i.\) \citet{survey} gives a rather complete survey of
applications of these homogeneous polynomials.
\\  \\ \noindent 
\textbf{\emph{Example 16:}} To get
\(B_{5,3}(y_1, y_2, y_{3})\) run

\begin{verbatim}
> eBellPol(5,3)
[1] 15(y1)(y2^2) + 10(y1^2)(y3)
\end{verbatim}

\noindent To get \(G_4(1, 1, 1, 1; y_1, y_2, y_3, y_4)\) run

\begin{verbatim}
> eBellPol(4)
[1] (y1^4) + 6(y1^2)(y2) + 3(y2^2) + 4(y1)(y3) + (y4)
\end{verbatim}

The
\href{https://www.rdocumentation.org/packages/kStatistics/versions/2.0/topics/oBellPol}{\tt oBellPol}
function returns the partial (ordinary) Bell polynomials
\[\hat{B}_{i,j}(y_1, \ldots, y_{i-j+1}) = \frac{j!}{i!} B_{i,j}(1! y_1, 2! y_2, \ldots, (i-j+1)! y_{i-j+1})\]
and the complete (ordinary) Bell polynomials
\[\hat{G}_i(y_1, \ldots, y_i) =  G_i(1, \ldots, 1; 1! y_1, 2! y_2, \ldots, i! y_i).\]

\hskip-0.5cm\textbf{\emph{Example 17:}} To get
\(\hat{B}_{5,3}(y_1, y_2,y_3)\) run

\begin{verbatim}
> oBellPol(5,3)
[1] 1/120( 360(y1)(y2^2) + 360(y1^2)(y3) )
\end{verbatim}

\noindent To get \(\hat{G}_3(y_1, y_2, y_3, y_4)\) run

\begin{verbatim}
> oBellPol(4)
[1] 1/24( 24(y1^4) + 72(y1^2)(y2) + 24(y2^2) + 48(y1)(y3) + 24(y4) )
\end{verbatim}

The
\href{https://www.rdocumentation.org/packages/kStatistics/versions/2.0/topics/e_eBellPol}{\tt e\_eBellPol}
function evaluates the exponential Bell polynomials when the
indeterminates are substituted with numerical values. In the following
some special sequence of numbers obtained using this procedure:
\[{\small \begin{array}{c|c|c} 
                           \hbox{ }              & \hbox{Sequence}                 &  \hbox{Bell polynomials} \\  \hline 
\hbox{Lah numbers}                             & \frac{i!}{j!} \begin{pmatrix}
i-1 \\j-1
\end{pmatrix} & B_{i,j} (1!, 2!, 3!, \ldots) \\ \hline 
\hbox{Stirling numbers of first kind}          & s(i,j)                          & B_{i,j} (0!, -1!, 2!, \ldots) \\ \hline  
\hbox{unsigned Stirling numbers of first kind} & |s(i,j)|                        & B_{i,j} (0!, 1!, 2!, \ldots) \\ \hline 
\hbox{Stirling numbers of second kind}         & S(i,j)                          & B_{i,j} (1, 1, 1, \ldots) \\ \hline 
\hbox{idempotent numbers}                      & \begin{pmatrix}
i \\j
\end{pmatrix} j^{i-j}           & B_{i,j} (1, 2, 3, \ldots) \\ \hline 
\hbox{Bell numbers}                            & B_i                             & \sum_{j=0}^i B_{i,j} (1, 1, 1, \ldots)
\end{array}}\] By default, the
\href{https://www.rdocumentation.org/packages/kStatistics/versions/2.0/topics/e_eBellPol}{\tt e\_eBellPol}
function returns the Stirling numbers of second kind, as the following
example shows.
\\  \\ \noindent 
\textbf{\emph{Example 18:}} To get \(S(5,3)\) run

\begin{verbatim}
> e_eBellPol(5,3)
[1] 25
> e_eBellPol(5,3,c(1,1,1,1,1))
[1] 25
\end{verbatim}

\noindent To get the \(5\)-th Bell number \(B_5\) run

\begin{verbatim}
> e_eBellPol(5)
[1] 52
\end{verbatim}

\noindent To get \(s(5,3)\) run

\begin{verbatim}
> e_eBellPol(5,3, c(1,-1,2,-6,24))
[1] 35
\end{verbatim}

\hypertarget{composition-of-formal-power-series}{%
\section{Composition of formal power
series}\label{composition-of-formal-power-series}}

In \(\eqref{mfaa1}\), suppose \(f_{\boldsymbol{t}}\) the
\(\boldsymbol{t}\)-th coefficient of \(f(\boldsymbol{x})\) and
\(g_{1; \boldsymbol{s}}, \ldots, g_{n; \boldsymbol{s}}\) the
\(\boldsymbol{s}\)-th coefficients of
\(g_1(\boldsymbol{z}), \ldots, g_n(\boldsymbol{z})\) respectively. Using
multi-index partitions, the multivariate Faà di Bruno's formula
\(\eqref{multfaa}\) can be written as \citep{MR2773373}
\begin{equation}\label{multfaa2}
h_{\boldsymbol{i}} = \boldsymbol{i}! \sum_{{\Lambda} \vdash \boldsymbol{s}_1, \ldots, \tilde{\Lambda} \vdash  \boldsymbol{s}_n \atop \boldsymbol{s}_1 + \cdots + \boldsymbol{s}_n = \boldsymbol{i}} f_{(l(\Lambda),\ldots,l(\tilde{\Lambda}))} \frac{g_{1, \Lambda} \cdots g_{n, \tilde{\Lambda}}}{\Lambda!
\cdots \tilde{\Lambda}! \, \mathfrak{m}(\Lambda)! \cdots \mathfrak{m}(\tilde{\Lambda})!}
\end{equation} where \(g_{1,\Lambda}, \ldots, g_{n,\tilde{\Lambda}}\)
are given in \(\eqref{sequences}\) and the sum is over all the
partitions
\({\Lambda} \vdash \boldsymbol{s}_1, \ldots, \tilde{\Lambda} \vdash \boldsymbol{s}_n,\)
with \((\boldsymbol{s}_1,\ldots,\boldsymbol{s}_n)\) \(m\)-tuples of
non-negative integers such that
\(\boldsymbol{s}_1 + \cdots + \boldsymbol{s}_n = \boldsymbol{i}.\)

The
\href{https://www.rdocumentation.org/packages/kStatistics/versions/2.0/topics/MFB}{\tt MFB}
function generates all the summands of \(\eqref{multfaa2}\). Its first
step is to find the set \(\tilde{p}(n,\boldsymbol{i})\) of all the
compositions of \(\boldsymbol{i}\) in \(n\) parts, that is all the
\(m\)-tuples \((\boldsymbol{s}_1,\ldots,\boldsymbol{s}_n)\) of
non-negative integers such that
\(\boldsymbol{s}_1 + \cdots + \boldsymbol{s}_n = \boldsymbol{i}.\) This
task is performed by the
\href{https://www.rdocumentation.org/packages/kStatistics/versions/2.0/topics/mkT}{\tt mkT}
function.

\noindent

\rule{13cm}{0.4pt}

\textbf{\emph{Function
\href{https://www.rdocumentation.org/packages/kStatistics/versions/2.0/topics/mkT}{mkT}}}

\begin{quote}
\emph{i)} Find all the partitions \(\Lambda \vdash \boldsymbol{i},\)
using the
\href{https://www.rdocumentation.org/packages/kStatistics/versions/2.0/topics/mkmSet}{\tt mkmSet}
function.
\end{quote}

\begin{quote}
\emph{ii)} Select the first partition \(\Lambda.\) If
\(l(\Lambda) = n,\) then the columns of \(\Lambda\) are the \(m\)-tuples
\((\boldsymbol{s}_1, \ldots, \boldsymbol{s}_n)\) such that
\(\boldsymbol{s}_1 + \ldots + \boldsymbol{s}_n = \boldsymbol{i}.\) If
\(l(\Lambda) < n,\) add \(n-l(\Lambda)\) zero columns to \(\Lambda.\)
\end{quote}

\begin{quote}
\emph{iii)} Generate all the permutations of the columns of \(\Lambda\)
as collected at step \emph{ii)}.
\end{quote}

\begin{quote}
\emph{iv)} Repeat steps \emph{ii)} and \emph{iii)} for each partition
\(\Lambda\) carried out at step \emph{i)}.
\end{quote}

\noindent

\rule{13cm}{0.4pt}

An input flag variable in the
\href{https://www.rdocumentation.org/packages/kStatistics/versions/2.0/topics/mkT}{\tt mkT}
function permits to obtain the output in a more compact set up. See the
following example.
\\  \\ \noindent 
\textbf{\emph{Example 19:}} Suppose we are looking for the
elements of the set \(\tilde{p}(2,(2,1)),\) that is the pairs
\((\boldsymbol{s}_1, \boldsymbol{s}_2)\) such that
\(\boldsymbol{s}_1 + \boldsymbol{s}_2 = (2,1).\) Consider the partitions
of \((2,1)\) as given in Example 2. Then run

\begin{verbatim}
> mkT(c(2,1),2,TRUE)
[( 0 1 )( 2 0 )]
[( 2 0 )( 0 1 )]
[( 1 0 )( 1 1 )]
[( 1 1 )( 1 0 )]
[( 2 1 )( 0 0 )]
[( 0 0 )( 2 1 )]
\end{verbatim}

Note that \texttt{{[}(\ 2\ 1\ )(\ 0\ 0\ ){]}} and
\texttt{{[}(\ 0\ 0\ )(\ 2\ 1\ ){]}} are obtained adding a zero column to
the partition \texttt{{[}(\ 2\ 1\ ),\ \ 1\ {]}}, and then permuting the
two columns. No zero columns are added to
\texttt{{[}(\ 2\ 0\ )(\ 0\ 1\ ){]}} as the length of the partition is
\(2.\) The same is true for \texttt{{[}(\ 0\ 1\ )(\ 2\ 0\ ){]}} or
\texttt{{[}(\ 1\ 1\ )(\ 1\ 0\ ){]}} which are only permuted.

The
\href{https://www.rdocumentation.org/packages/kStatistics/versions/2.0/topics/MFB}{\tt MFB}
function produces the multivariate Faà di Bruno's formula
\(\eqref{multfaa2}\) making use of the following steps.

\noindent

\rule{13cm}{0.4pt}

\textbf{\emph{Function
\href{https://www.rdocumentation.org/packages/kStatistics/versions/2.0/topics/MFB}{\tt MFB}}}

\begin{quote}
\emph{i)} Find all the \(m\)-tuples
\((\boldsymbol{s}_1,\ldots,\boldsymbol{s}_n)\) in
\(\tilde{p}(n,\boldsymbol{i})\) using the
\href{https://www.rdocumentation.org/packages/kStatistics/versions/2.0/topics/mkT}{\tt mkT}
function.
\end{quote}

\begin{quote}
\emph{ii)} Let \(y_1, \ldots, y_n\) be variables. For each
\(j=1, \ldots, n,\) compute all the partitions
\(\Lambda \vdash \boldsymbol{s}_j\) using the
\href{https://www.rdocumentation.org/packages/kStatistics/versions/2.0/topics/mkmSet}{\tt mkmSet}
function and find the explicit expression of the polynomial \[
q_{j,\boldsymbol{s}_j}(y_j) = \boldsymbol{\boldsymbol{s}_j}! \sum_{{\Lambda} \vdash \boldsymbol{s}_j} y_j^{l(\Lambda)} \frac{g_{j, \Lambda}}{\Lambda!\mathfrak{m}(\Lambda)!}.
\]
\end{quote}

\begin{quote}
\emph{iii)} Make the products
\(q_{1,\boldsymbol{s}_1}(y_1) \cdots q_{n,\boldsymbol{s}_n}(y_n)\) in
the multivariable polynomial
\[{\small h_{\boldsymbol{i}}(y_1, \ldots, y_n)=\sum_{(\boldsymbol{s}_1,\ldots,\boldsymbol{s}_n) \in \tilde{p}(n,\boldsymbol{i})} {\boldsymbol{i} \choose \boldsymbol{s}_1,\ldots,\boldsymbol{s}_n} q_{1,\boldsymbol{s}_1}(y_1) \cdots q_{n,\boldsymbol{s}_n}(y_n)}\]
and compute its explicit expression.
\end{quote}

\begin{quote}
\emph{iv)} In the explicit expression of the polynomial
\(h_{\boldsymbol{i}}(y_1, \ldots, y_n)\) as carried out at the previous
step \emph{iii)}, replace the occurrences of the products
\(y_1^{l(\Lambda)} \cdots y_n^{l(\tilde{\Lambda})}\) with
\(f_{(l(\Lambda),\ldots,l(\tilde{\Lambda}))}.\)
\end{quote}

\noindent

\rule{13cm}{0.4pt}

Step \emph{iii)} is performed by the
\href{https://www.rdocumentation.org/packages/kStatistics/versions/2.0/topics/joint}{\tt joint}
function. This function is not directly accessible in the package, as
defined locally in the
\href{https://www.rdocumentation.org/packages/kStatistics/versions/2.0/topics/MFB}{\tt MFB}
function. The
\href{https://www.rdocumentation.org/packages/kStatistics/versions/2.0/topics/joint}{\tt joint}
function realizes a recursive pair matching: each coefficient
\(g_{1, \Lambda}\) of \(q_{1,\boldsymbol{s}_1}(y_1)\) is matched with
each coefficient \(g_{2, \tilde{\Lambda}}\) of
\(q_{2,\boldsymbol{s}_2}(y_2),\) then each paired coefficient
\(g_{1, \Lambda} g_{2, \tilde{\Lambda}}\) is matched with each
coefficient \(g_{3, \Lambda^{\!*}}\) of \(q_{3,\boldsymbol{s}_3}(y_3)\)
and so on. Step \emph{iv)} consists of multiplying each coefficient
found at step \emph{iii)} with \(f_{\boldsymbol{t}},\) where
\(\boldsymbol{t}\) is the multi-index whose \(j\)-th component gives how
many times \(g_{j, \cdot}\) appears in this coefficient. See the
following example.
\\  \\ \noindent 
\textbf{\emph{Example 20:}} Suppose \(n=m=2\) and
\(\boldsymbol{i}=(1,1).\) To get \(h_{(1,1)}\) in \(\eqref{multfaa2}\)
run

\begin{verbatim}
> MFB(c(1,1),2)
[1] f[1,1]g1[0,1]g2[1,0] + f[1,1]g1[1,0]g2[0,1] + f[2,0]g1[0,1]g1[1,0] + 
f[1,0]g1[1,1] + f[0,2]g2[0,1]g2[1,0] + f[0,1]g2[1,1]
\end{verbatim}

Taking into account \(\eqref{multps1}\), in the previous output
\texttt{f{[}i,j{]}} corresponds to \(f_{(i,j)}\) as well as
\texttt{g1{[}i,j{]}} and \texttt{g2{[}i,j{]}} correspond to
\(g_{1;(i,j)}\) and \(g_{2;(i,j)}\) respectively for \(i,j=0,1,2.\) Note
that \texttt{g1{[}1,1{]}} is multiplied with \texttt{f{[}1,0{]}} as
there is one occurrence of \texttt{g1} and no occurrence of \texttt{g2}.
In the same way, \texttt{g1{[}1,0{]}g1{[}0,1{]}} is multiplied with
\texttt{f{[}2,0{]}} as there are two occurrences of \texttt{g1} and no
occurrence of \texttt{g2} and \texttt{g1{[}1,0{]}g2{[}0,1{]}} is
multiplied with \texttt{f{[}1,1{]}} as there is one occurrence of
\texttt{g1} and one occurrence of \texttt{g2} and so on. Compare the
previous output with the one obtained in \texttt{Maple} running

\begin{quote}
diff(f(g1(x1,x2),g2(x1,x2),x1,x2): \[
\begin{array}{l}
D_{{2,2}}(f)(g1(x1,x2), g2(x1,x2))  \left(\!{\frac {\partial }{\partial {\it x1}}}{\it g2}(x1,x2)\!\right)\!\! \left(\!{\frac {\partial }{\partial x2}} g2(x1,x2)\!\right) \\ + D_{{1,2}}(f)(g1(x1,x2),g2(x1,x2)) \left(\!{\frac {\partial }{\partial x2}}g1(x1,x2)\!\right)\!\! 
\left(\!{\frac {\partial }{\partial x1}}g2(x1,x2)\!\right) \\
+ D_{1,2}(f)(g1(x1,x2),g2(x1,x2)) \!\! 
\left(\!{\frac {\partial }{\partial x1}}g1(x1,x2))\!\right) \!\!
\left(\!{\frac {\partial }{\partial x2}}g2(x1,x2)\!\right) \\
+ D_{{1,1}}(f)(g1(x1,x2), g2(x1,x2))
\left(\! {\frac {\partial}{\partial x1}}g1(x1,x2))\!\right)\!\!
\left(\!{\frac {\partial }{\partial x2}}g1(x1,x2))\!\right) \\
+ D_{{2}}(f)(g1(x1,x2), g2(x1,x2)) 
\left({\frac {\partial^{2}}{\partial x2 \partial x1}}g2(x1,x2)\right)\\
+ D_{{1}}(f)(g1(x1,x2),g2(x1,x2)) \left({\frac {\partial^{2}}{\partial x2 \partial x1}} g1(x1,x2)\right)
\end{array}\] where \(D_{1}(f)\) denotes
\(\partial f(x_1,x_2)/\partial x_1, D_{2}(f)\) denotes
\(\partial f(x_1,x_2)/\partial x_2\) and similarly
\[{\small D_{1,1}(f) \leftarrow  \frac{\partial^2 f(x_1,x_2)}{\partial x_1^2}, \, D_{2,2}(f) \leftarrow  \frac{\partial^2 f(x_1,x_2)}{\partial x_2^2}, \, D_{1,2}(f) \leftarrow \frac{\partial^2 f(x_1,x_2)}{\partial x_1 \partial x_2}}.\]
\end{quote}

The
\href{https://www.rdocumentation.org/packages/kStatistics/versions/2.0/topics/eMFB}{\tt eMFB}
function evaluates the multivariate Faà di Bruno's formula
\(\eqref{multfaa2}\) when the coefficients of the formal power series
\(f\) and \(g_1, \ldots, g_n\) in \(\eqref{multps1}\) are substituted
with numerical values.
\\  \\ \noindent 
\textbf{\emph{Example 21:}} To evaluate the output of
Example 20 for some numerical values of the coefficients, run

\begin{verbatim}
> cfVal<-"f[0,1]=2, f[0,2]=5, f[1,0]=13, f[1,1]=-4, f[2,0]=0"
> cgVal<-"g1[0,1]=-2.1, g1[1,0]=2,g1[1,1]=3.1,g2[0,1]=5,g2[1,0]=0,g2[1,1]=6.1"
> cVal<-paste0(cfVal,",",cgVal)
> e_MFB(c(1,1),2,cVal)
[1] 12.5
\end{verbatim}

The polynomial families discussed in the previous sections are generated
using the
\href{https://www.rdocumentation.org/packages/kStatistics/versions/2.0/topics/MFB}{\tt MFB}
function. Indeed, the generalized (complete exponential) Bell
polynomials in \(\eqref{(sol22ter)}\) are coefficients of the following
formal power series \begin{equation}
\hskip-1cm{\small H(y_1, \ldots, y_n; \boldsymbol{z}) = 1 + \sum_{|\boldsymbol{i}| > 0} h_{\boldsymbol{i}}(y_1, \ldots, y_n) \frac{{\boldsymbol{z}}^{\boldsymbol{i}}}{\boldsymbol{i}!} = \exp \bigg[ \sum_{i=1}^n y_i (g_i(\boldsymbol{z})-1) \bigg]},
\label{(GCBellH)}
\end{equation} which turns to be a composition \(\eqref{mfaa1}\), with
\(f(x_1, \ldots, x_n) =\) \(\exp(x_1 y_1 + \cdots + x_n y_n)\) and
\(f_{\boldsymbol{t}} = y_1^{t_1} \cdots y_n^{t_n}\) for
\(\boldsymbol{t} \in {\mathbb N}_0^n.\) In this case,
\(y_1, \ldots, y_n\) play the role of indeterminates. The
\(\boldsymbol{i}\)-th coefficient
\(h_{\boldsymbol{i}}(y_1, \ldots, y_n)\) - output of the
\href{https://www.rdocumentation.org/packages/kStatistics/versions/2.0/topics/GCBellPol}{\tt GCBellPol}
function - is obtained from the multivariate Faà di Bruno's formula
\(\eqref{multfaa2}\) dealing with \(y_1, \ldots, y_n\) as they were
constants. When \(\{g_1(\boldsymbol{z}), \ldots, g_n(\boldsymbol{z})\}\)
are the same formal power series \(g(\boldsymbol{z}),\) the formal power
series \(H(y_1, \ldots, y_n;\boldsymbol{z})\) in \(\eqref{(GCBellH)}\)
reduces to \begin{equation}
{\small H(y_1, \ldots, y_n; \boldsymbol{z}) = \exp \big[ (y_1 + \cdots + y_n) (g(\boldsymbol{z})-1) \big]}
\label{Hequal}
\end{equation} with coefficients
\(\tilde{h}_{\boldsymbol{i}}(y_1, \ldots, y_n)\) as given in the
previous section.

If \(n=1\) then \(H(y_1, \ldots, y_n; \boldsymbol{z})\) reduces to the
composition \(\exp \big[ y (g(\boldsymbol{z})-1)]\) whose coefficients
are the polynomials given in \(\eqref{(redGC)}\). More in general the
coefficients of \(f(g(\boldsymbol{z})-1)\) are \begin{equation}
h_{\boldsymbol{i}} =  \sum_{{\Lambda} \vdash  \boldsymbol{i}} d_{\Lambda}  f_{l(\Lambda)} 
g_{\Lambda}
\label{redGC1}
\end{equation} where the sum is over all the partitions
\(\Lambda=(\boldsymbol{\lambda}_1^{r_1} , \boldsymbol{\lambda}_2^{r_2}, \ldots) \vdash \boldsymbol{i},\)
\(d_{\Lambda}\) is given in \(\eqref{dlm}\) and
\(g_{\Lambda} = g_{\boldsymbol{\lambda}_1}^{r_1} g_{\boldsymbol{\lambda}_2}^{r_2} \cdots.\)
If also \(m=1,\) then \(h_{\boldsymbol{i}}\) in \(\eqref{redGC1}\)
reduces to \begin{equation}
h_{i} =  \sum_{{\lambda} \vdash  i} d_{\lambda}  f_{l(\lambda)} 
g_{\lambda}
\label{redGC2}
\end{equation} where the sum is over all the partitions
\(\lambda=(1^{r_1} , 2^{r_2}, \ldots) \vdash i,\) \(d_{\lambda}\) is
given in \(\eqref{dlambda}\) and
\(g_{\lambda} = g_1^{r_1} g_{2}^{r_2} \cdots.\) Formula
\(\eqref{redGC2}\) corresponds to the univariate Faà di Bruno's formula
and gives the \(i\)-th coefficient of \(f(g(z)-1)\) with
\[f(x)=1+\sum_{j \geq 1} f_j \frac{x^j}{j!} \qquad \hbox{and} \qquad g(z)=1+\sum_{s \geq 1} g_s \frac{z^s}{s!}.\]
\\  \\ \noindent 
\textbf{\emph{Example 22:}} To get \(h_{5}\) in \(\eqref{redGC2}\) run

\begin{verbatim}
>MFB(c(5), 1)
[1] f[5]g[1]^5 + 10f[4]g[1]^3g[2] + 15f[3]g[1]g[2]^2 + 10f[3]g[1]^2g[3] + 
10f[2]g[2]g[3] + 5f[2]g[1]g[4] + f[1]g[5]
\end{verbatim}

For instance, the \(i\)-th general partition polynomial in
\(\eqref{(ffaa)}\) is generated using the
\href{https://www.rdocumentation.org/packages/kStatistics/versions/2.0/topics/MFB}{\tt MFB}
function: in such a case the univariate Faà di Bruno's formula
\(\eqref{redGC2}\) is generated with \(\{y_s\}\) in place of \(\{g_s\}\)
and \(\{a_j\}\) in place of \(\{f_j\}.\)

\hypertarget{examples-of-how-to-generate-polynomials-not-included-in-the-kstatistics-package}{%
\subsubsection{\texorpdfstring{Examples of how to generate polynomials
not included in the
\href{https://cran.r-project.org/web/packages/kStatistics/index.html}{\tt kStatistics}
package}{Examples of how to generate polynomials not included in the kStatistics package}}\label{examples-of-how-to-generate-polynomials-not-included-in-the-kstatistics-package}}

In the following we give some suggestions on how to use the \texttt{R}
codes of the
\href{https://cran.r-project.org/web/packages/kStatistics/index.html}{\tt kStatistics}
package to generate additional polynomial families.

The
\href{https://www.rdocumentation.org/packages/kStatistics/versions/2.0/topics/pPart}{\tt pPart}
function is an example of how to use the univariate Faà di Bruno's
formula and a symbolic strategy different from those presented so far.
Indeed the
\href{https://www.rdocumentation.org/packages/kStatistics/versions/2.0/topics/pPart}{\tt pPart}
function generates the so-called partition polynomial \(F_i(y)\) of
degree \(i,\) whose coefficients are the number of partitions of \(i\)
with \(j\) parts for \(j=1, \ldots, i\) \citep{MR2427666}. The partition
polynomial \(F_i(y)\) is obtained from the univariate Faà di Bruno's
formula \(\eqref{redGC2}\) setting \begin{equation}
f_j = 1/i! \qquad \hbox{and} \qquad g_{s}^{r_s} =(s!)^{r_s} r_s! y^{r_s}
\label{ppart}
\end{equation} for \(s=1, \ldots, i-j+1, j=1,\ldots,i\) and
\(r_s=1, \ldots, i.\) Note the symbolic substitution of \(g_{s}^{r_s}\)
with the powers \(y^{r_s}.\)
\\  \\ \noindent 
\textbf{\emph{Example 23:}} To get \(F_5(y)\) run

\begin{verbatim}
> pPart(5)
[1] y^5 + y^4 + 2y^3 + 2y^2 + y
\end{verbatim}

\noindent Note that \(F_5(y)\) is obtained from the output of Example 22
using \(\eqref{ppart}.\)
\\  \\ \noindent 
\textbf{\emph{Example 24:}} The following code shows how to
evaluate \(F_{11}(y)\) in \(y=7.\)

\begin{verbatim}
> s<-pPart(11)         # generate the partition polynomial of degree 11
> s<-paste0("1",s)     # add the coefficient to the first term  
> s<-gsub(" y","1y",s) # replace the variable y without coefficient  
> s<-gsub("y", "*7",s) # assign y = 7
> eval(parse(text=s))  # evaluation of the expression
[1] 3.476775e+182
\end{verbatim}

We give a further example on how to generate a polynomial family not
introduced so far but still coming from \(\eqref{redGC2}\) for suitable
choices of \(\{f_j\}\) and \(\{g_{s}\}.\) Consider the elementary
symmetric polynomials in the indeterminates \(y_1, \ldots, y_n\)
\begin{eqnarray}  
e_i(y_1, \ldots, y_n) = \left\{ \begin{array}{cl} 
\displaystyle{\sum_{1\leq j_1 < \cdots < j_i \leq n}} y_{j_1} \cdots y_{j_i}, & 1 \leq i \leq n, \\
0,      &  i > n.
\end{array}
\right.
\end{eqnarray} A well-known result \citep{MR1937238} states that these
polynomials can be expressed in terms of the power sum symmetric
polynomials \(\eqref{ps}\) in the same indeterminates
\(y_1, \ldots, y_n,\) using the general partition polynomials
\(\eqref{gpp}\), that is \begin{equation}
e_i = \frac{(-1)^i}{i!} G_i (1, \ldots, 1; - p_1, - 1! p_2, - 2! p_3, \ldots, -(i-1)! p_i)
\label{elps1}
\end{equation} for \(i=1, \ldots, n.\) The following \texttt{e2p}
function expresses the \(i\)-th elementary symmetric polynomial \(e_i\)
in terms of the power sum symmetric polynomials \(p_1, \ldots, p_i,\)
using \(\eqref{elps1}\) and the
\href{https://www.rdocumentation.org/packages/kStatistics/versions/2.0/topics/MFB}{\tt MFB}
function.

\begin{verbatim}
> e2p <-  function(n=0){
+     v<-MFB(n,1);                      # Call the MFB Function
+     v<-MFB2Set( v );                  # Expression to vector
+     for (j in 1:length(v)) {
+         # ----- read -----------[ fix block ]-----------------------#
+         c <- as.character(v[[j]][2]);    # coefficient
+         x <- v[[j]][3];                  # variable
+         i <- v[[j]][4];                  # subscript
+         k <- strtoi(v[[j]][5]);          # power
+         # ----- change --------------------------------------------#
+         if (x=="f") {
+             c<-paste0(c,"*( (-1)^",n,")");
+             x<-"";
+             i<-"";
+         }
+         else if (x=="g") {
+             c<-paste0(c,"*((-factorial(",strtoi(i)-1,"))^",k,")");
+             x<-paste0("(p",i,ifelse(k>1,paste0("^",k),""),")");
+             i<-"";k<-1;
+         }
+         # ----- write ---------[ fix block ]-----------------------#
+         v[[j]][2] <- c;
+         v[[j]][3] <- x;
+         v[[j]][4] <- i;
+         v[[j]][5] <- k;
+         # ---------------------------------------------------------#
+     }
+     noquote(paste0("1/",factorial(n),"( ",Set2expr(v), " )"));
+ }
\end{verbatim}

This function starts by initializing the vector \texttt{v} with
\(\eqref{redGC2}\) by means of the
\href{https://www.rdocumentation.org/packages/kStatistics/versions/2.0/topics/MFB}{\tt MFB}
function. There is a first code snippet \texttt{{[}fix\ block{]}} for
extracting a set with the coefficients, variables, indexes and powers of
\texttt{v} by means of the
\href{https://www.rdocumentation.org/packages/kStatistics/versions/2.0/topics/MFB2Set}{\tt MFB2Set}
function. This first code snippet should not be changed whatever
polynomial family we are generating. The second code snippet
\texttt{change} includes instructions that can be changed according to
the expressions of the coefficients \(\{f_j\}\) and \(\{g_{s}\}\) in
\(\eqref{redGC2}\). To get \(\eqref{elps1}\), we set \(f_j=1\) and
\(g_{s} = - (s-1)! p_s.\) Once these coefficients have been changed, the
last code snippet \texttt{{[}fix\ block{]}} updates the vector
\texttt{v}. The
\href{https://www.rdocumentation.org/packages/kStatistics/versions/2.0/topics/Set2expr}{\tt Set2expr}
function assembles the final expression.
\\  \\ \noindent 
\textbf{\emph{Example 25:}} To get \(e_4\) in
\(\eqref{elps1}\) run

\begin{verbatim}
> e2p(4)
[1] 1/24( (p1^4) - 6(p1^2)(p2) + 3(p2^2) + 8(p1)(p3) - 6(p4) )
\end{verbatim}

\hypertarget{concluding-remarks}{%
\section{Concluding remarks}\label{concluding-remarks}}

We have developed the
\href{https://cran.r-project.org/web/packages/kStatistics/index.html}{\tt kStatistics}
package with the aim to generate univariate and multivariate
\(k\)-statistics/polykays, togheter with the multivariate Faà di Bruno's
formula and various user-friendly functions related to this formula. The
paper briefly introduces the combinatorial tools involved in the package
and presents, in detail, the core function of the package which
generates multi-index partitions. We emphasize that the algorithms
presented here have been designed with the aid of the umbral calculus,
even if we did not mentioned this method in the paper.

One of the main applications we have dealt with is the generation and
evaluation of various families of polynomials: from generalized complete
Bell polynomials to general partition polynomials, from partial Bell
polynomials to complete Bell polynomials. Numerical sequences obtained
from the Bell polynomials can also be generated.

All these utilities intend to make the
\href{https://cran.r-project.org/web/packages/kStatistics/index.html}{\tt kStatistics}
package a useful tool not only for statisticians but also for users who
need to work with families of polynomials usually available in symbolic
software or tables. Indeed, we have provided examples on how to generate
polynomial families not included in the package but which can still be
recovered using the Faà di Bruno's formula and suitable strategies, both
numerical and symbolic. Following this approach, also the estimations of
joint cumulants or products of joint cumulants is one further example of
symbolic strategy coming from the multivariate Faà di Bruno's formula.

Future works consist in expanding the
\href{https://cran.r-project.org/web/packages/kStatistics/index.html}{\tt kStatistics}
package by including extensions of the multivariate Faà di Bruno's
formula, as addressed in \citet{MR2186441} and references therein,
aiming to manage nested compositions, as the \texttt{BellY} function in
the Wolfram Language and System does. Moreover, further procedures can
be inserted relied on symbolic strategies not apparently related to the
multivariate Faà di Bruno's formula but referable to this formula, as
for example the central Bell polynomials \citep{sym11020288}.

The results in this paper were obtained using the
\href{https://cran.r-project.org/web/packages/kStatistics/index.html}{\tt kStatistics}
\(2.1.1\) package. The package is currently available with a general
public license (GPL) from the Comprehensive \texttt{R} Archive Network.

\bibliographystyle{apalike}
\bibliography{RJarxiv}     

\end{document}